\documentclass[conference]{IEEEtran}
\IEEEoverridecommandlockouts

\usepackage{amssymb, multirow}

\usepackage{cite}

\usepackage{amssymb,amsfonts,amsthm}
\usepackage[fleqn]{amsmath}

\usepackage{algorithmic}
\usepackage{graphicx}
\usepackage{textcomp}
\usepackage{xcolor}

\usepackage{algorithm}

\usepackage{graphicx}
\usepackage{caption}
\usepackage{subcaption}
\usepackage{url}

\usepackage{balance}

\usepackage{mdframed}

\def\BibTeX{{\rm B\kern-.05em{\sc i\kern-.025em b}\kern-.08em
    T\kern-.1667em\lower.7ex\hbox{E}\kern-.125emX}}

\usepackage{todonotes}

\DeclareCaptionLabelFormat{andtable}{#1~#2  \&  \tablename~\thetable}

\begin{document}

\title{Adaptive Deep Learning of Cross-Domain Loss in Collaborative Filtering}

\author{\IEEEauthorblockN{Dimitrios Rafailidis}
\IEEEauthorblockA{\textit{Maastricht University} \\
Maastricht, The Netherlands \\
dimitrios.rafailidis@maastrichtuniversity.nl}

\and

\IEEEauthorblockN{Gerhard Weiss}
\IEEEauthorblockA{\textit{Maastricht University} \\
Maastricht, The Netherlands \\
gerhard.weiss@maastrichtuniversity.nl}
}

\maketitle

\begin{abstract}
Nowadays, users open multiple accounts on social media platforms and e-commerce sites, expressing their personal preferences on different domains. However, users' behaviors change across domains, depending on the content that users interact with, such as movies, music, clothing and retail products. In this paper, we propose an adaptive deep learning strategy for cross-domain recommendation, referred to as ADC. We design a neural architecture and formulate a cross-domain loss function, to compute the non-linearity in user preferences across domains and transfer the knowledge of users' multiple behaviors, accordingly. In addition, we introduce an efficient algorithm for cross-domain loss balancing which directly tunes gradient magnitudes and adapts the learning rates based on the domains' complexities/scales when training the model via backpropagation. In doing so, ADC controls and adjusts the contribution of each domain when optimizing the model parameters. Our experiments on six publicly available cross-domain recommendation tasks demonstrate the effectiveness of the proposed ADC model over other state-of-the-art methods. Furthermore, we study the effect of the proposed adaptive deep learning strategy and show that ADC can well balance the impact of the domains with different complexities. 
\end{abstract}


\section{Introduction}\label{sec:intro}

With the advent of social media platforms and e-commerce systems, such as Amazon and Netflix, users express their preferences in multiple domains~\cite{AliannejadiRC17,RafailidisC16,RafailidisC17}. For example, users can rate items from different domains, such as books and retail products on Amazon, or users express their opinion on different social media platforms, such as Facebook and Twitter. The sparsity of user preferences in each domain limits the recommendation accuracy of collaborative filtering strategies~\cite{CREM10,Her04,Kor09}. To overcome the shortcomings of single-domain models, cross-domain recommendation strategies exploit\footnote{Different from multimedia retrieval~\cite{RafailidisNM11} and cross-domain retrieval~\cite{sigir/RafailidisC16}, in this study cross-domain recommendation aims at capturing multi-preferences across different platforms.} the additional information of user preferences in multiple auxiliary/source domains to face the sparsity problem and leverage the recommendation accuracy in a target domain~\cite{LI09}. The source domains can be categorized based on users' and items' overlaps, that is, full-overlap, and partial or non user/item overlap between the domains~\cite{CREM11,TKDD15}. In this study, we focus on partial users' overlaps between the target and the source domains, as it reflects on the real-world setting, where for instance users open multiple accounts on social media platforms or interact with different types of content on the same platform~\cite{Khan17}. Generating cross-domain recommendations is a challenging task, as the source domains might be a potential source of noise, for example, if user preferences differ in the multiple domains, then the source domains are likely to introduce noise in the learning of the target domain~\cite{BER07,CREM11,GAO13,PAN10}. Therefore, cross-domain recommendation strategies aim to learn how to transfer the knowledge of user preferences from different domains by weighting the importance of users' different behaviors accordingly.

In the relevant literature cross-domain recommendation strategies, such as~\cite{GAO13,HU13,LI09,Lon14}, form user and item clusters to capture the relationships between multiple domains at a cluster level, thus tackling the sparsity problem; and then weigh the user preferences to generate the top-$N$ recommendations in the target domain. However, such cross-domain strategies linearly combine the cluster-based  user preferences in the target domain, which does not reflect on the real-world world scenario with users having complex behaviors across domains. More recently, to capture the non-linear associations of users' different preferences in multiple domains a few deep learning architectures have been designed for cross-domain recommendation~\cite{EL15,Hu18,Wang19}. Although these cross-domain deep learning strategies try to learn the non-linearity in users' multi-aspect preferences across domains, they do not account for the different complexities/scales of multiple domains, which might reduce the quality of recommendations. For example, if the source domains are richer than the target domain, existing deep learning cross-domain recommendation algorithms learn how to recommend items in the source domains and consider the target domain as noise. Therefore, a pressing challenge resides on how to adapt the learning process of neural architectures in cross-domain deep learning strategies, while considering the different complexities/scales of multiple domains. In particular, deep learning strategies of user preferences in various domains need to be properly balanced so that the neural network's parameters converge to robust shared parameters that are useful across all domains. 

Recently, various deep learning strategies have been introduced to perform balanced learning of different tasks with various complexities/scales in natural language processing~\cite{Has17}, speech synthesis~\cite{Wu15}, image segmentation~\cite{Heg17} and object detection~\cite{Red17}. Multi-task deep learning strategies aim to find this balance by controlling the forward pass in a neural network e.g., by either constructing explicit statistical relationships between features or optimizing multi-task network architectures~\cite{Ken18,Mis16}. However, these neural models do not capture users' multiple preferences across domains, and ignore the complexity/scale of user preferences in a domain that might be too dominant when optimizing the model parameters.

In this paper, we propose an Adaptive Deep Learning model for Cross-domain recommendation, namely ADC. In particular our contribution is summarized as follows:  

\begin{itemize}
\item We design a neural architecture and formulate a cross-domain loss function to capture the non-linear associations between user preferences across different domains.
\item We propose an efficient algorithm for cross-domain loss balancing which directly tunes gradient magnitudes and adapts the learning rates based on the domains' complexities when training the model. In doing so, ADC controls and adjusts the contribution of each domain when optimizing the model parameters.
\end{itemize} In our experiments on six cross-domain recommendation tasks from Amazon, we show the superiority of the proposed ADC model over other baseline methods, demonstrating the effectiveness of the proposed adaptive neural learning strategy.

The remainder of the paper is organized as follows, Section~\ref{sec:rel} reviews related work of cross-domain recommendation, and then Section~\ref{sec:prop} presents the proposed ADC model. Finally, in Section~\ref{sec:exp} we evaluate the performance of the proposed model against other baseline models, and Section~\ref{sec:conc} concludes the study.

\section{Related Work}\label{sec:rel}
Cross-domain recommendation algorithms differ in how the perform knowledge transfer of user preferences across domains to produce recommendations~\cite{Khan17}. For example, the graph-based method of~\cite{CREM11} models the similarity relationships as a direct graph and explore all possible paths connecting users or items to capture the cross-domain relationships. Hu et al.~\cite{HU13} model a cubic user-item-domain matrix (tensor), and by applying factorization the respective latent space is constructed to generate cross-domain recommendations. Li et al.~\cite{LI09} compute user and item clusters for each domain, and then encode the cluster-based patterns in a shared codebook. Finally, the knowledge of user preferences is transferred across domains through the shared codebook. Gao et al.~\cite{GAO13} calculate the latent factors of user-clusters and item-clusters to construct a common latent space, which represents the preference patterns e.g., rating patterns, of user clusters on the item clusters. Then, the common cluster-based preference pattern that is shared across domains is learned following a subspace strategy, so as to control the optimal level of sharing among multiple domains. Cross-domain collaborative filtering with factorization machines (FMs), presented in~\cite{Lon14}, is a state-of-the-art cross-domain recommendation which extends FMs~\cite{Rendle12}. It is a context-aware approach which applies factorization on the merged domains, aligned by the shared users, where the source domains are used as context.  However, these cross-domain recommendation strategies do not capture the non-linearity of users' preferences in multiple domains~\cite{Hu18}.

To overcome the limitations of linear cluster-based strategies, Elkahky et al.~\cite{EL15} propose a multi-view deep learning approach for content-based recommendation across domains. In their model, the item and user features are acquired from different sources of Microsoft products, such as Windows Apps recommendation, News recommendation, and Movie/TV recommendation. Wang et al.~\cite{Wang19} introduce a multi-task feature learning algorithm for knowledge graph enhanced recommendation, introducing a deep end-to-end framework that utilizes a knowledge graph embedding task for the recommendation task. The two tasks are associated by cross and compress units, which automatically share latent features and learn high-order interactions between items in the recommendation system and entities in the knowledge graph. Hu et al.~\cite{Hu18} jointly learn neural networks to generate cross-domain recommendations based on stich units~\cite{Mis16}, introducing a shared auxiliary matrix to couple two hidden layers when training the networks in parallel. However, these deep learning strategies for cross-domain recommendation ignore the domains' different complexities during learning and do not weigh the loss functions accordingly. Consequently, the non-weighting of domains' loss functions might degrade the quality of recommendations as we will show in our experiments in Section~\ref{sec:exp}.

\section{The Proposed ADC Model} \label{sec:prop}
\subsection{Cross-domain Loss Function} \label{sec:cross}
In our setting we assume that we have $p$ different domains, where $n_k$ and $m_k$ are the numbers of users and items in the $k$-th domain, respectively, with $k=1,\dots,p$. In each matrix $R^{(k)}$, we store the user-item interactions  e.g., ratings, number of clicks or views, and so on. 
We define two disjoint sets, a set ${\mathcal{I}^+_u}^{(k)}$ of observed items that user $u$ has already interacted with in domain $k$, and a set ${\mathcal{I}^-_u}^{(k)}$ of  unobserved items. For each observed item $i^+\in {\mathcal{I}^+_u}^{(k)}$, we randomly sample negative/unobserved items $i^-\in {\mathcal{I}^-_u}^{(k)}$, for each user $u$. According to the Bayesian Pairwise Ranking (BPR) criterion~\cite{BPR}, the goal is to learn to rank the observed items higher than the unobserved ones, having the following single-domain loss function for each domain $k$:
\begin{equation} \label{eq:loss}
L_k = - \sum_{(u, i^+,i^-)} \log \sigma({\mathbf{u}^{(k)}_u}^\top {\mathbf{v}^{(k)}_{i+}} - {\mathbf{u}^{(k)}_u}^\top  {\mathbf{v}^{(k)}_{i-}})
\end{equation}
where $\sigma(x)=1/\big(1+\exp(-x)\big)$ is the logistic sigmoid function. For each domain $k$ we consider the user latent vectors ${\mathbf{u}^{(k)}_u} \in \mathbb{R}^{d \times 1}$ and item latent vectors ${\mathbf{v}^{(k)}_i} \in \mathbb{R}^{d \times 1}$, with $u=1,\ldots,n_k$,  $i=1,\ldots,m_k$ and $d$ being the number of latent dimensions.

In this study we consider users' partial overlaps across the $p$ domains. This means that at the same time we have to regularize the different user latent vectors of the same user $u$ across all the domains. Notice that a user might not exist in all the $p$ domains. In case that user $u$ does not appear in a $k$-th domain we average her latent vector by all the latent vectors of the domains that she expressed her preferences. Thus, we reformulate the single-domain loss function of Equation~\ref{eq:loss} as follows:

\begin{multline} \label{eq:loss2}
L_k = - \sum_{(u, i^+,i^-)} \log \sigma({\mathbf{u}^{(k)}_u}^\top {\mathbf{v}^{(k)}_{i+}} - {\mathbf{u}^{(k)}_u}^\top  {\mathbf{v}^{(k)}_{i-}})  \\ +   \lambda\sum\limits_{q \neq k} || \mathbf{u}_u^{(q)} - \mathbf{u}_u^{(k)} ||_2^2
\end{multline}
where the last term expresses the approximation error of the user latent vectors in domains $q$ and $k$, and $\lambda$ controls the regularization. Provided that we have $p$ different domains, in our problem we have to jointly learn all the $p$ different single-domain loss functions of Equation~\ref{eq:loss2}, formulating the following \emph{cross-domain loss function}: 
\begin{equation} \label{eq:loss3}
L_{cross} = \sum_{k=1}^{p}w_k L_k
\end{equation}

\textbf{Weights' Adaptive Learning.} Weights $w_k$ control the influence of each single-domain loss function when optimizing the joint cross-domain loss function. In our problem we have to perform adaptive learning of weights $\{w_1,w_2,\ldots,w_p \}$ while training our model based on the cross-domain loss function in Equation~\ref{eq:loss3}. As we will show in Section~\ref{sec:arc} a weight $w_k$  directly couples to the backpropagated gradient magnitudes for each domain, thus the challenge is to find the optimal value for each $w_k$ at each learning step that balances the contribution of each domain to the learning process of the cross-domain loss function. To optimize the weights $w_k$, we propose an adaptive learning algorithm in our model's cross-domain neural network that penalizes the network when backpropagated gradients from any domain are too large or too small (Section~\ref{sec:adap}). The correct balance is achieved when domains are learned at similar rates; if a domain is learned relatively quickly, then its weight $w_k$ should decrease relative to other domain weights to allow other domains to more influence the model training. In particular, when learning the weights of the cross-domain loss function we aim to the following goals:

\begin{itemize}
\item Provided that the $p$ different domains might have different complexity, we have to adaptively learn the weight $w_k$ of each domain in each iteration/epoch during the optimization. To achieve this, we have to compute the gradients of the weights accordingly, by placing gradient norms for the $p$ different domains on a common scale based on their relative magnitudes.
\item In addition, we have to dynamically adjust gradient norms so that the $p$ different single-domain loss functions in Equation~\ref{eq:loss3} are minimized at similar learning rates.
\end{itemize}

\subsection{ADC Neural Architecture}\label{sec:arc}
\begin{figure*} [t]
  \centering
    \includegraphics[width=0.55\textwidth]{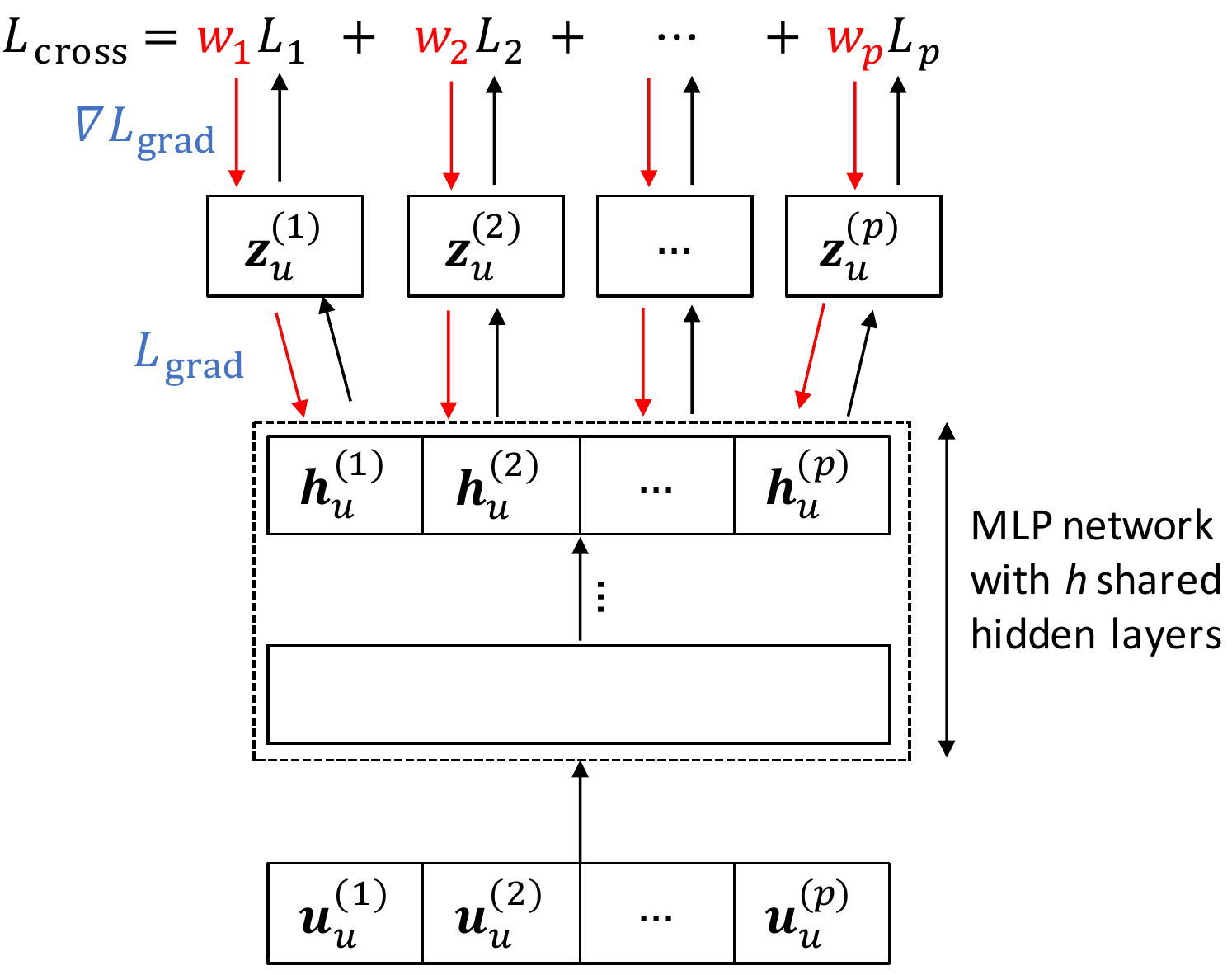}
\caption{An overview of the proposed ADC model. The MLP network with $h$ hidden layers computes the non-linear associations between the user latent vectors in the $p$ domains. The decoupled hidden reprentations $\mathbf{z}_u^{(k)}$ are learned based on the weights in a matrix $W$ using a loss function $L_{grad}$. Finally, the weights $w_k$ of the cross-domain loss function $L_{cross}$ are adjusted based on the gradients $\nabla L_{grad}$ in each backpropagation step.} \label{fig:over}    
\end{figure*}

Figure~\ref{fig:over} presents an overview of the proposed cross-domain neural network, which consists of the following components:

\textbf{(1):} The input layer at the bottom of our architecture is a concatenated vector $\mathbf{u}_{con}$ of the $p$ latent vectors of user $u$. To initialize the user latent vectors we factorize  each matrix $R^{(k)}$ separately, that is minimizing the approximation error $\sum_{(u,i)} || R^{(k)}_{ui} - {\mathbf{u}^{(k)}_u}^\top {\mathbf{v}^{(k)}_{i}} ||^2$ for each domain $k=1,\ldots,p$.

\textbf{(2):} The MLP network with the $h$ hidden layers is a fully-connected network with ReLU activations. The MLP network tries to learn the non-linear associations between the user latent vectors. This is achieved by computing the MLP network's weights $\mathcal{A}$ via standard gradient descent in the backward pass of the backpropagation algorithm.

\textbf{(3):} The hidden representations of the ($h$-th) last shared layer of the MLP network is split/decoupled into $p$ different $d$-dimensional representations $\mathbf{z}_u^{(k)}$, which are called \emph{decoupled hidden representations} for each domain $k$ and are computed based on the weights in matrix $W$. In particular, matrix $W$ is the last shared layer of weights of the decoupled hidden representations where we perform adaptive learning and apply normalization to gradients based on a loss function $L_{grad}$ to consider the different complexities of each domain $k$, as we will describe in Section~\ref{sec:adap}.  

\textbf{(4):} Having computed the gradients $\nabla w_k L_{grad}$,  $\forall k=1,\ldots,p$, we update/adjust the weights $w_k$ of the cross-domain loss function $L_{cross}$.

\subsection{Adaptive Learning in the Cross-domain Neural Network}\label{sec:adap}
\textbf{Gradient Norms \& Learning Rates.}  In our proposed neural architecture we consider the following gradient norms and learning rates:
\begin{itemize}
\item $G_W^{(k)}(t) = || \nabla_W w_k(t)L_k(t) ||_2$: the $L2$ norm of the gradient of the weighted single-domain loss $w_k(t)L_k(t)$ with respect to the weights in $W$.
\item $\hat{G}_W(t)=E_{dom}[G_W^{(k)}(t)]$: the average gradient norm across all the $p$ domains at iteration $t$.
\item $\tilde{L}_k(t)=L_k(t)/L_k(0)$: the single-domain loss learning ratio for domain $k$ at iteration $t$. Lower values of $\tilde{L}_k(t)$ correspond to a faster learning rate for domain $k$. In our implementation we set $L_k(0) = \log{(p)}$.
\item $r_k(t)=\tilde{L}_k(t)/E_{dom}[\tilde{L}_k(t)]$: the relative inverse learning rate of the $k$-th domain.
\end{itemize}

\textbf{The Loss Function $L_{grad}$.} At the $t$-th iteration of our algorithm the cross-domain loss function is formulated as follows,  $L_{cross}(t) = \sum_{k=1}^{p}w_k(t) L_k(t)$. As stated in Section~\ref{sec:cross} our model has to compute a common scale for gradient magnitudes, and control the learning rates of the different domains at the last layer of the decoupled hidden representations. The common scale for gradients is the average gradient norm $\hat{G}_W(t)$ which serves as a baseline at each iteration $t$ to determine the relative gradient sizes. The relative inverse learning rate of a domain $r_k(t)$ is used to rate balance the gradients of all $p$ domains. The higher the value of $r_k(t)$, the higher the gradient magnitudes should be for domain $k$, forcing the respective single-domain loss function $L_k(t)$ to be learned more quickly. Therefore, the gradient norm for the $k$-th domain is computed as follows:
\begin{equation}\label{eq:tun}
G_W^{(k)}(t) \leftarrow \hat{G}_W(t) [r_k(t)]^\gamma , \quad \forall k=1,\ldots,p
\end{equation}
We define a hyperparameter $\gamma$ as the \emph{asymmetry parameter} of our model which controls the strength of forcing the single-domain loss functions $L_k$ to be learned at a common learning rate. In cases where domains are very different in their complexity, leading to significantly different learning rates between the domains over the optimization algorithm, a higher value of $\gamma$ should be used to enforce stronger learning rate balancing. On the other hand, when domains are more symmetric a lower value of $\gamma$ is required. Note that $\gamma$=0 will set the norms of the backpropagated gradients from each domain to be equal to $W$. The effects of $\gamma$ on the model's performance is further studied in Section~\ref{sec:param} .

Equation~\ref{eq:tun} gives a target norm for each $k$-th domain's gradient norm. In our algorithm we have to update our loss weights $w_k(t)$ to move the gradient norm towards the target norm for each domain. We define an $L1$ loss function $L_{grad}$ between the actual and target gradient norms at each iteration for each domain, summed over all domains as follows:
\begin{equation}\label{eq:lgrad}
\min_{w_k} L_{grad} = \sum_{k=1}^p || G_W^{(k)}(t) - \hat{G}_W(t) [r_k(t)]^\gamma ||_1
\end{equation}
When differentiating the loss $L_{grad}$ with respect to the weight $w_k$, we treat the target gradient norm $\hat{G}_W(t) [r_k(t)]^\gamma$ as a fixed constant, with $w_k$ controlling the gradient magnitude per domain. The computed gradients $\nabla w_k L_{grad}$ are then used to update each $w_k$ via gradient descent as shown in Figure~\ref{fig:over}. 

Our adaptive learning algorithm of the proposed ADC model in the cross-domain neural network is outlined as follows. Having computed the user and item latent vectors for each domain separately, then we initialize the weight $w_k$ of the cross-domain loss function $L_{cross}$. To initialize the MLP network's weights in $\mathcal{A}$, we first train our model with random initializations using only one hidden layer in the MLP network. Then, the trained parameters are used to initialize the MLP network's weights for $h$ hidden layers. Finally, we split our training data into mini-batches of concatenated vectors $\mathbf{u}_{con}$, and in each iteration we train our model to adjust the weights $w_k$ while minimizing the cross-domain loss function $L_{cross}$, as described in Section~\ref{sec:arc}.

\section{Experimental Evaluation} \label{sec:exp}
\subsection{Cross-domain Recommendation Tasks} Our experiments were performed on six cross-domain tasks from the Amazon dataset\footnote{\url{https://snap.stanford.edu/data/}}~\cite{Les07}. The items are grouped in categories/domains, and we evaluate the performance of our model on the six largest domains. The main characteristics of the evaluation data are presented in Table~\ref{tab:data}, showing the sparsity in user preferences for each domain. Notice that the domains have different complexities/scales, corresponding to the real-world setting. For example, the largest domain ``music'' contains 174K ratings of 69K users on 24K items, whereas the smallest domain ``toys'' has 13K ratings of 9K users on 3K items.

\begin{table}[h]
\centering
\caption{The six cross-domain recommendation tasks.} \label{tab:data}
\vspace{-0.2cm}
\begin{center}%
\begin{tabular}{l@{\quad\quad}c@{\quad}c@{\quad}c@{\quad}c}\hline
Domain & Users & Items & Ratings & Density (\%) \\ \hline
electronics & 18,649 & 3,975  &  23,009 & 0.031\\
kitchen & 16,114 & 5,511  & 19,856 & 0.022\\
toys & 9,924 & 3,451 & 13,147 & 0.038\\
dvd & 49,151 & 14,608 & 124,438 & 0.017\\
music & 69,409 & 24,159 & 174,180  & 0.010\\
video & 11,569  &5,223  & 36,180  & 0.059\\ \hline
\end{tabular}
\end{center}
\end{table}

\subsection{Evaluation Protocol} In each out of the six cross-domain recommendation tasks, the goal is to generate recommendations for a target domain, while the remaining five domains are considered as source domains. We trained the examined models on the 50\% of the ratings of the six domains. For each cross-domain recommendation task, we used 10\% of the ratings in the target domain as cross-validation set to tune the models' parameters and evaluate the examined models on the remaining test ratings. To remove user rating bias from our results, we considered an item as relevant if a user has rated it above her average ratings and irrelevant otherwise~\cite{Rec17}. We measured the quality of the top-$N$ recommendations in terms of the ranking-based metrics recall and Normalized Discounted Cumulative Gain (NDCG@N). Recall is the ratio of the relevant items in the top-$N$ ranked list over all the relevant items for each user. NDCG measures the ranking of the relevant items in the top-$N$ list. For each user the Discounted Cumulative Gain (DCG) is defined as: $$DCG@N = \sum_{j=1}^{N}{\frac{2^{rel_j}-1}{\log_2{j+1}}}$$ where $rel_j$ represents the relevance score of item $j$, that is binary in our case, i.e., relevant or irrelevant. NDCG is the ratio of DCG/iDCG, where iDCG is the ideal DCG value given the ratings in the test set. We repeated our experiments five times and in our results we report average recall and NDCG over the five runs. 

\subsection{Compared Methods}
In our experiments we compare the following methods:
\begin{itemize}
\item \emph{\textbf{BPR}}~\cite{BPR}: a baseline ranking model that tries to rank the observed/rated items over the unobserved ones. BPR is a single-domain method and does not exploit users' preferences across different domains when generating recommendations. 
\item \emph{\textbf{CLFM}}~\cite{GAO13}: a cross-domain Cluster-based Latent Factor Model which uses joint non-negative tri-factorization to construct a latent space to represent the rating patterns of user clusters on the item clusters from each domain, and then generates the cross-domain recommendations based on a subspace learning strategy.
\item \emph{\textbf{CDCF}}~\cite{Lon14}: a cross-domain collaborative filtering strategy that extends factorization machines (FMs) by factorizing the merged domains, aligned by the shared users.
\item \emph{\textbf{ScoNet}}~\cite{Hu18}: a cross-domain model that jointly learns stich networks, with a shared auxiliary matrix to couple two hidden layers when training the networks in parallel. In our experiments, we used the variant of ScoNet with $L1$-norm to force the matrices to be sparse, as suggested in~\cite{Hu18}.
\item \emph{\textbf{ADC}}: the proposed cross-domain model that performs adaptive weighting of the cross-domain loss function, by adjusting the learning rates and gradient magnitudes, to control the influence of each domain when training the model based on the domains' complexities.
\end{itemize}

\begin{table}[t]
\centering
\caption{Performance evaluation in terms of NDCG@10. Bold values denote the best scores, using the paired t-test ($p<$0.05). } \label{tab:res1}
\vspace{-0.2cm}
\begin{center}%

\begin{tabular}{l@{\quad}c@{\quad}c@{\quad}c@{\quad}c@{\quad}c@{\quad}c}\hline
			 	& BPR  & CLFM & CDCF & ScoNet & ADC & Improv. (\%)\\\hline
electronics 		&	0.1938 &	0.2385 &	0.2577 &	0.2791 &	\textbf{0.3037} & \textbf{8.81}  \\
kitchen 			&	0.1356	 & 0.1502 &	0.1589 &	0.1774 &	\textbf{0.1952} & \textbf{10.03}\\
toys			&	0.1624	 & 0.1763 &	 0.1894 & 0.2013 & \textbf{0.2195} & \textbf{9.04}\\
dvd 		&	.0.3847	& 0.4012 &	0.4457 &	0.4662 &	\textbf{0.4980} & \textbf{6.82}\\
music 	&	0.3291	& 0.3728	& 0.4032	& 0.4322	 & \textbf{0.4748}	&  \textbf{9.85}\\  
video 			&	 0.5421 &	0.5711 &	0.6336 &	0.6805 &	\textbf{0.7073} & \textbf{3.93} \\ \hline
\end{tabular}
\end{center}
\end{table}

\begin{figure*}[t]
\centering
\begin{tabular}{cc}  
\hspace{-0.5cm}\includegraphics[width=\columnwidth]{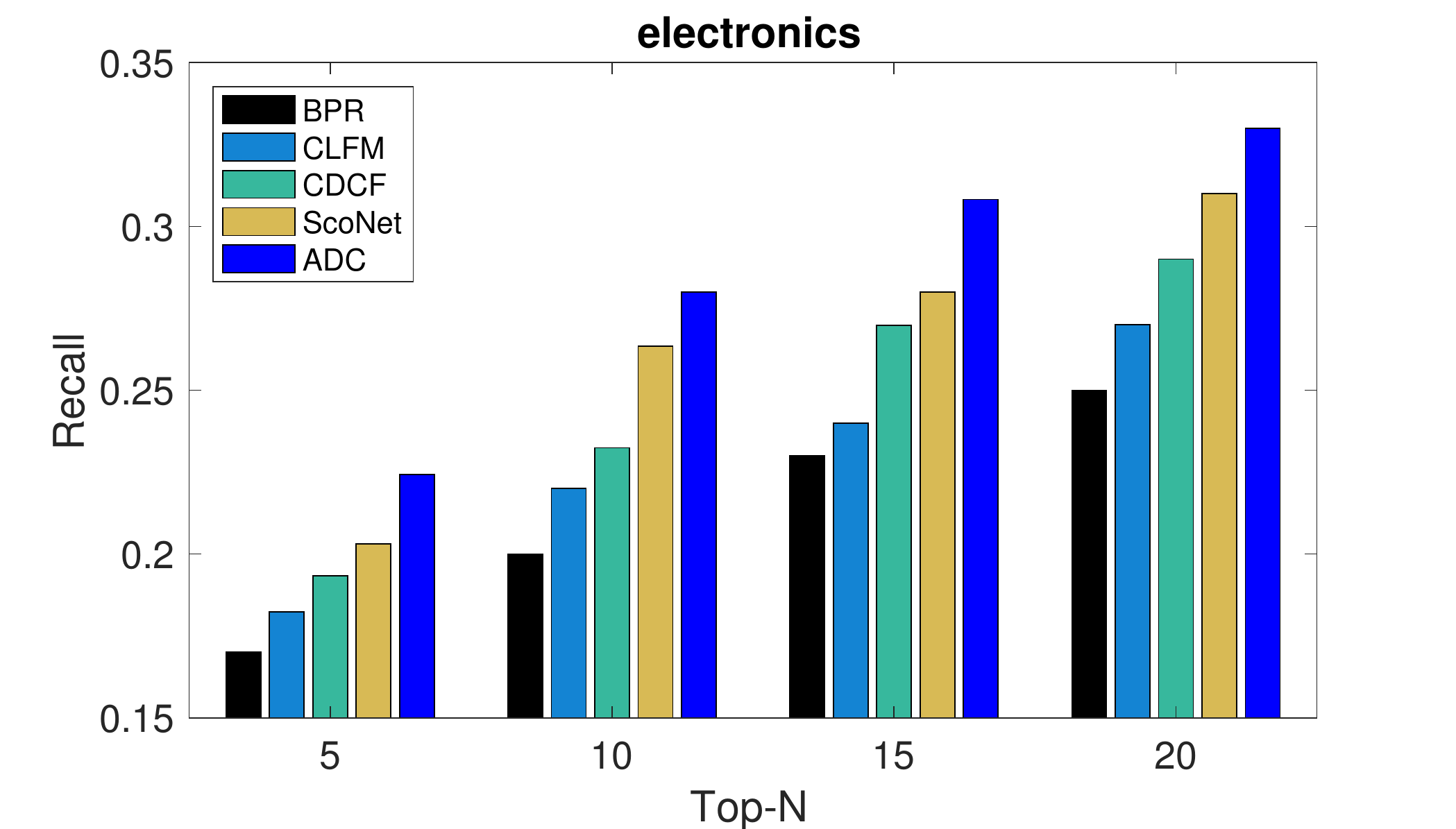} & \hspace{-0.75cm}\includegraphics[width=\columnwidth]{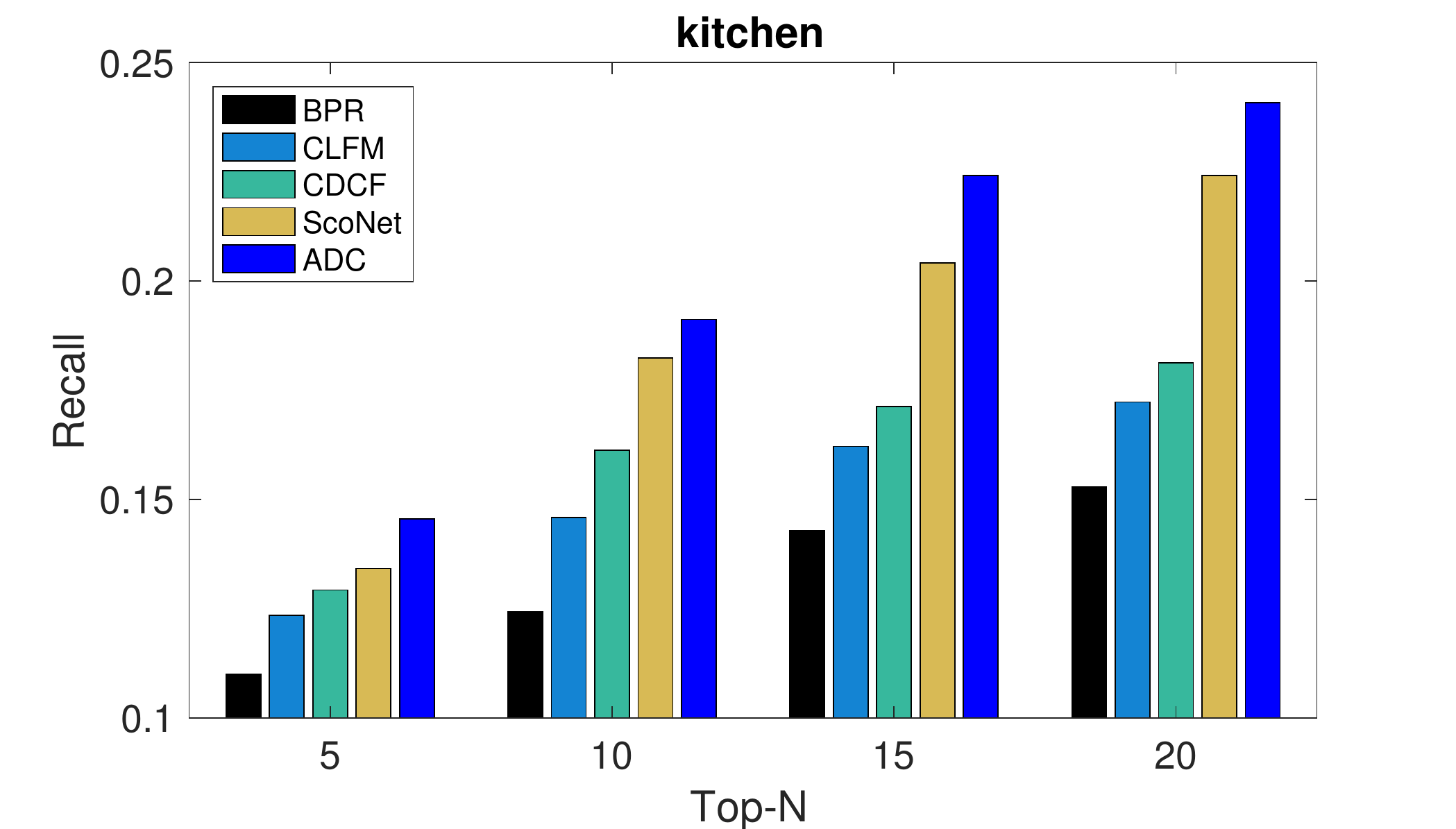}\\ 
\hspace{-0.5cm}\includegraphics[width=\columnwidth]{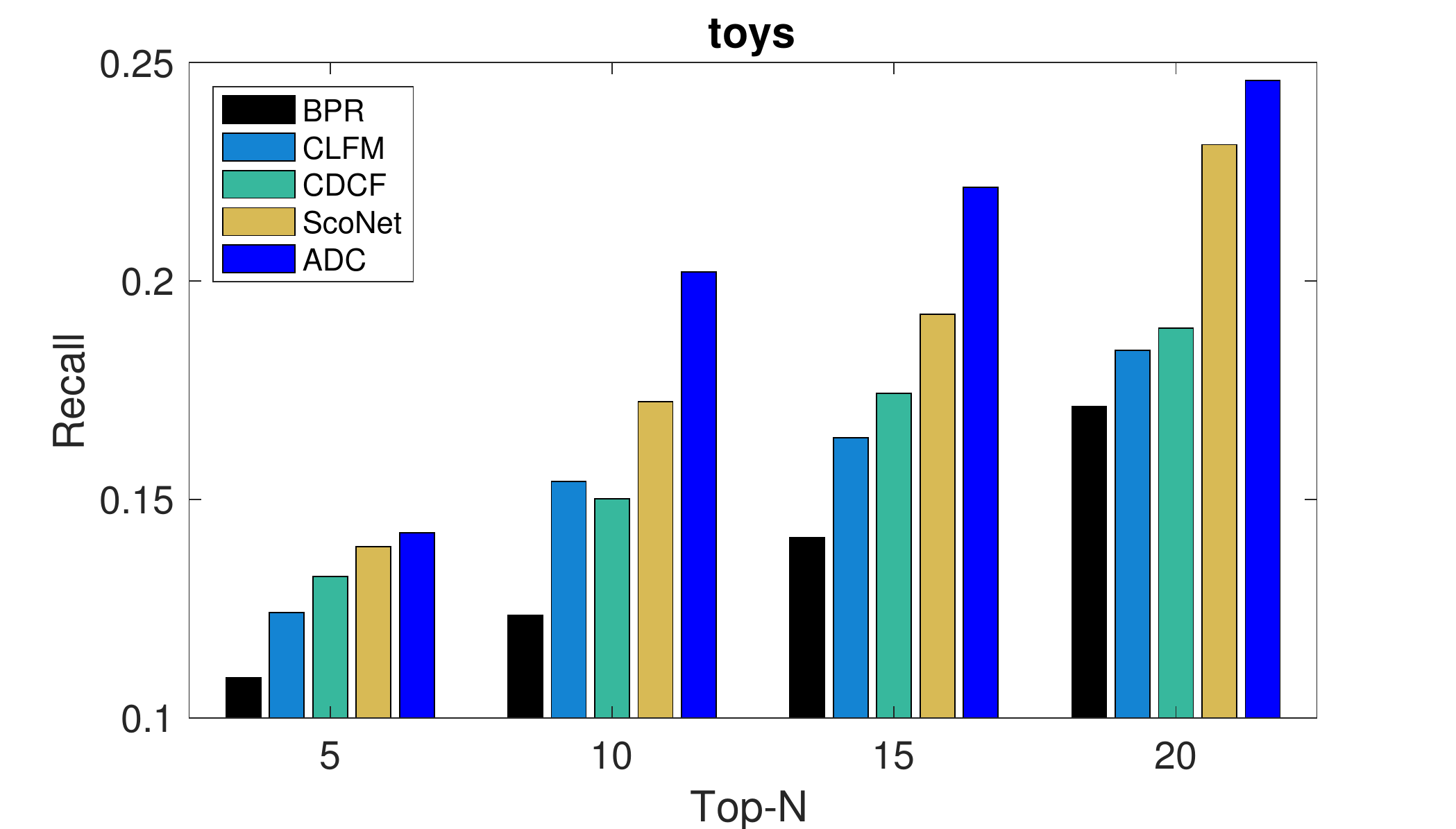} & \hspace{-0.75cm} \includegraphics[width=\columnwidth]{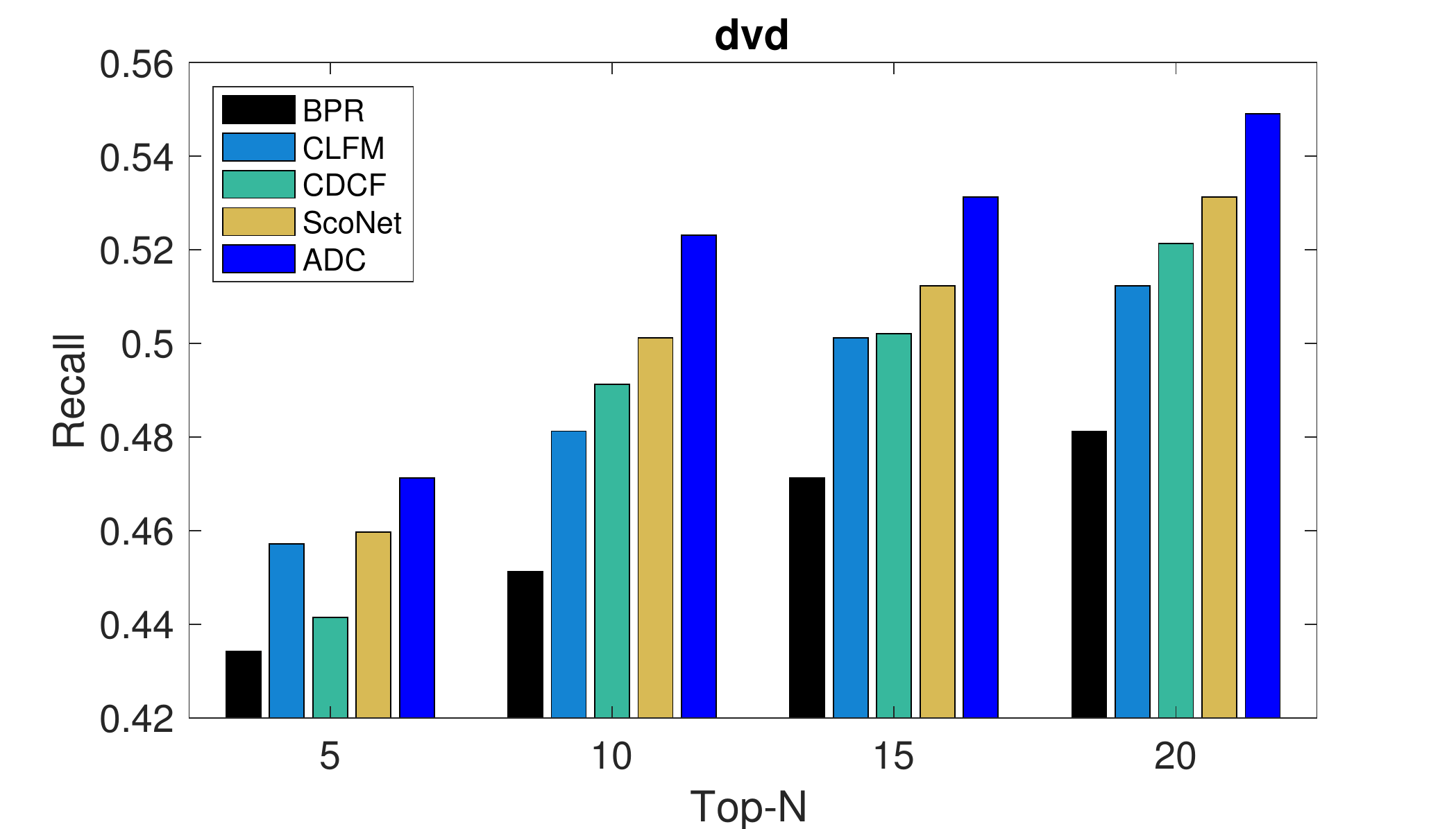} \\
\hspace{-0.5cm}\includegraphics[width=\columnwidth]{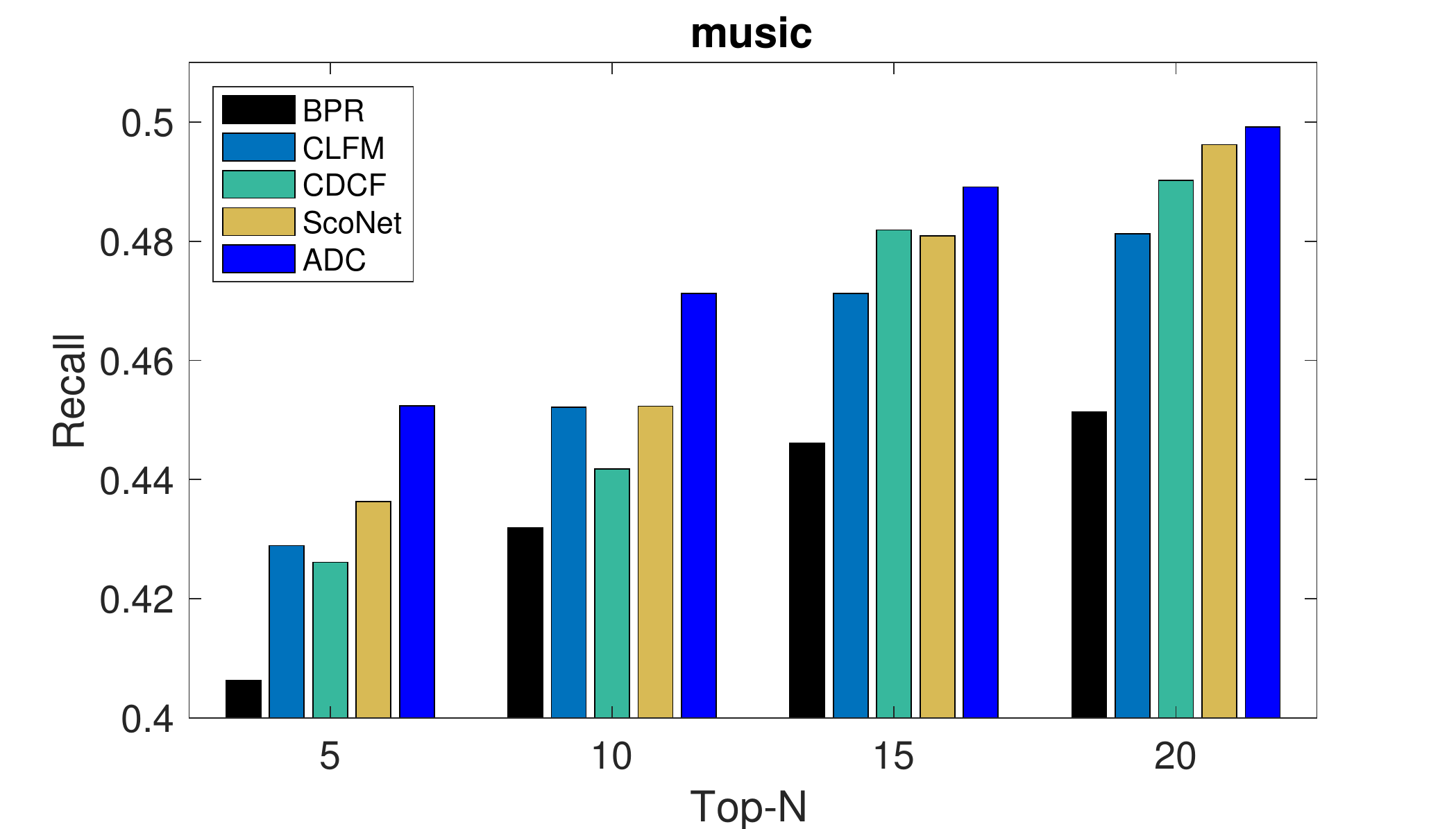} & \hspace{-0.75cm}\includegraphics[width=\columnwidth]{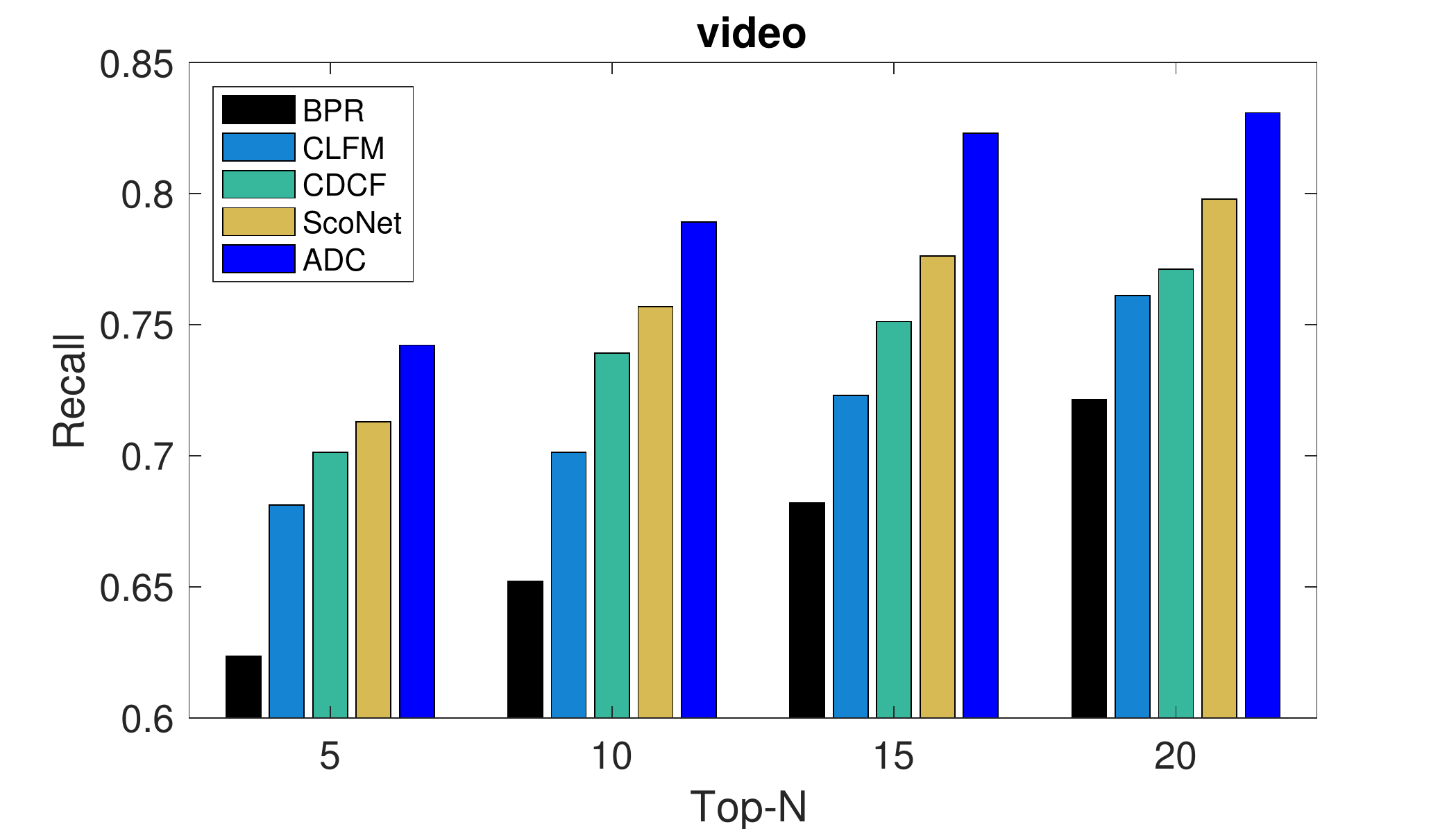}\\ 
\end{tabular}
\vspace{-0.1cm}\caption{Effect on recall when varying the top-$N$ recommendations.} \label{fig:res1}
\end{figure*}

\subsection{Performance Evaluation}
Table~\ref{tab:res1} presents the performance of the examined models in terms of NDCG@10. In addition, Figure~\ref{fig:res1} reports the experimental results in terms of recall when varying the number of top-$N$ recommendation in 5, 10, 15 and 20. On inspection of the results, we observe that all the cross-domain recommendation strategies outperform the single-domain BPR strategy in all cases, indicating the crucial role of incorporating the knowledge of user preferences from multiple domain when generating recommendations. Compared to the cluster-based strategies CLFM and CDCF, the most competitive deep learning model ScoNet performs better, as ScoNet can capture the non-linearity in user preferences across the six domains. Using the paired t-test we found that ADC is superior over all the competitive approaches for $p<$0.05. The proposed ADC model achieves an average improvement of 6.14 and 8.08\% in term of recall and NDCG respectively, when compared with the second best method ScoNet. This occurs because ADC not only can compute the non-linear associations among users' different preferences in the six domains based on the cross-domain loss function of our neural architecture, but also ADC adjusts the weights of each single-domain loss function and controls the contribution of each domain when training the models based on the domains' complexities.

\begin{figure*}[t] 
\centering
\begin{tabular}{cc} 
\includegraphics[width=\columnwidth]{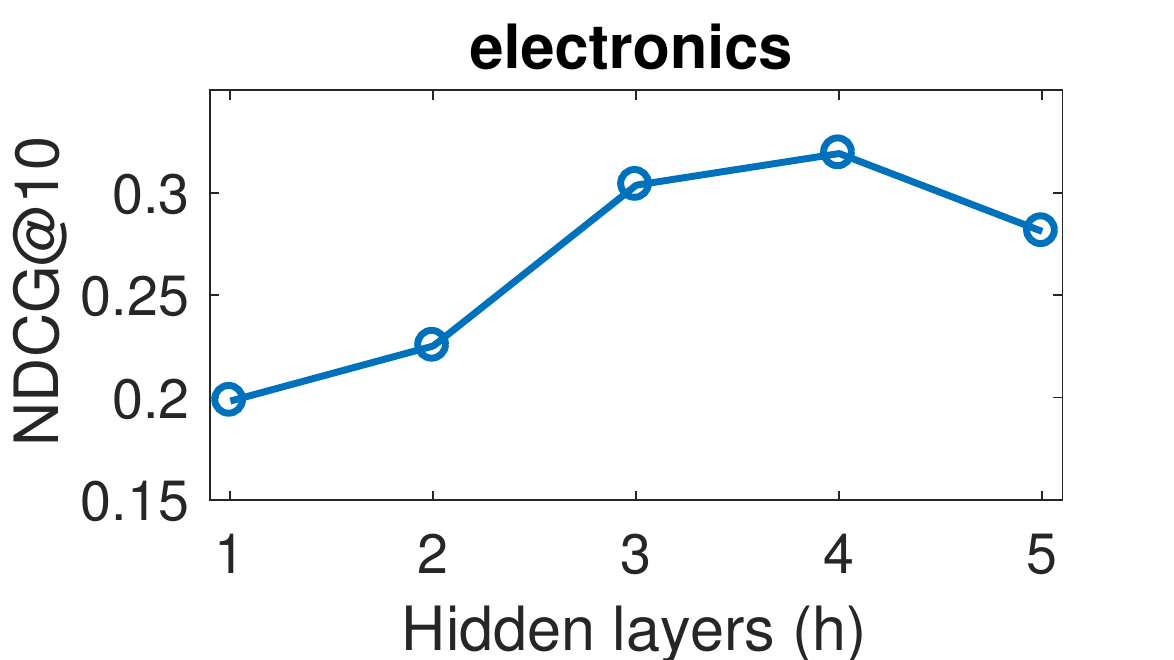} \hspace{0.1cm} & \includegraphics[width=\columnwidth]{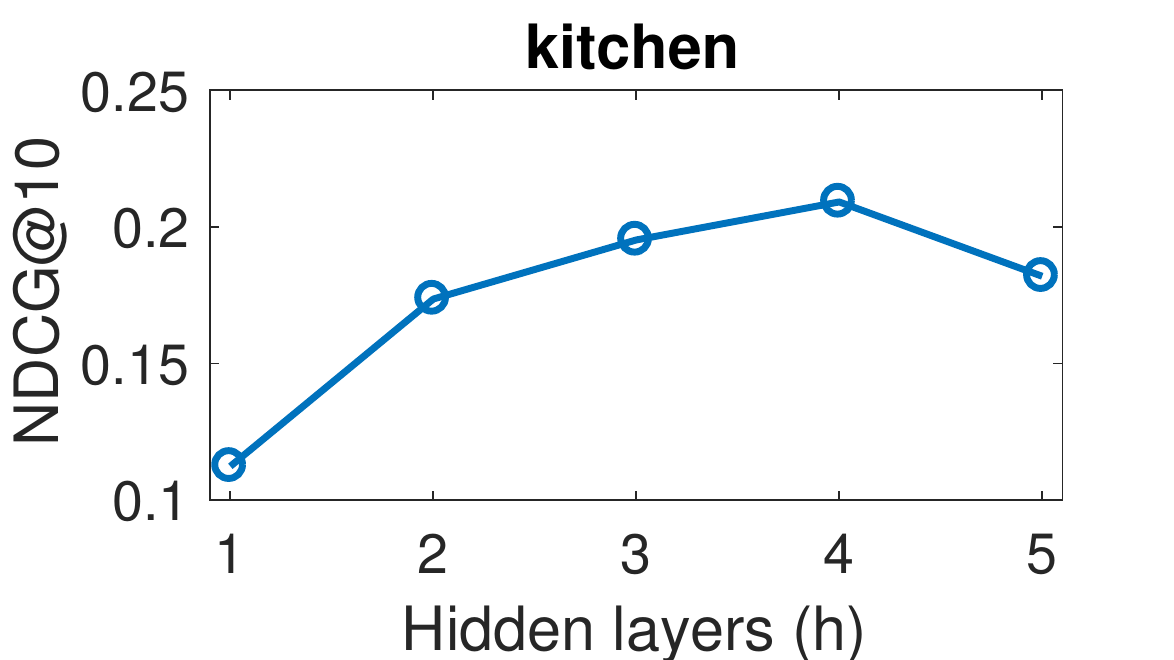} \\
\includegraphics[width=\columnwidth]{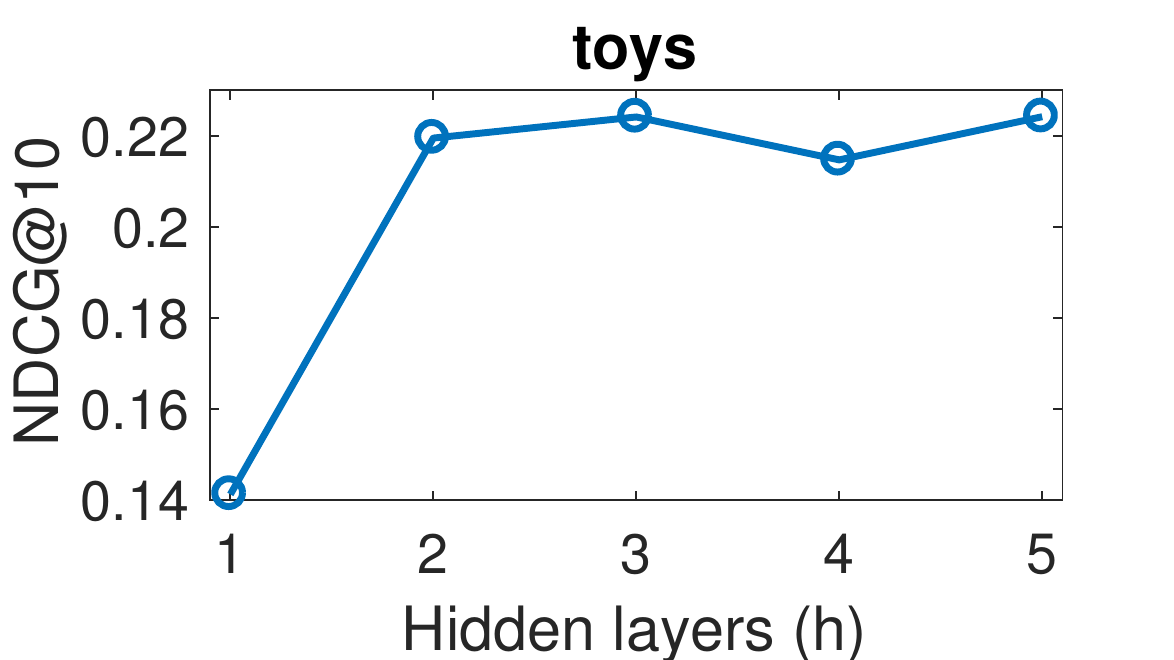}  \hspace{0.1cm} & \includegraphics[width=\columnwidth]{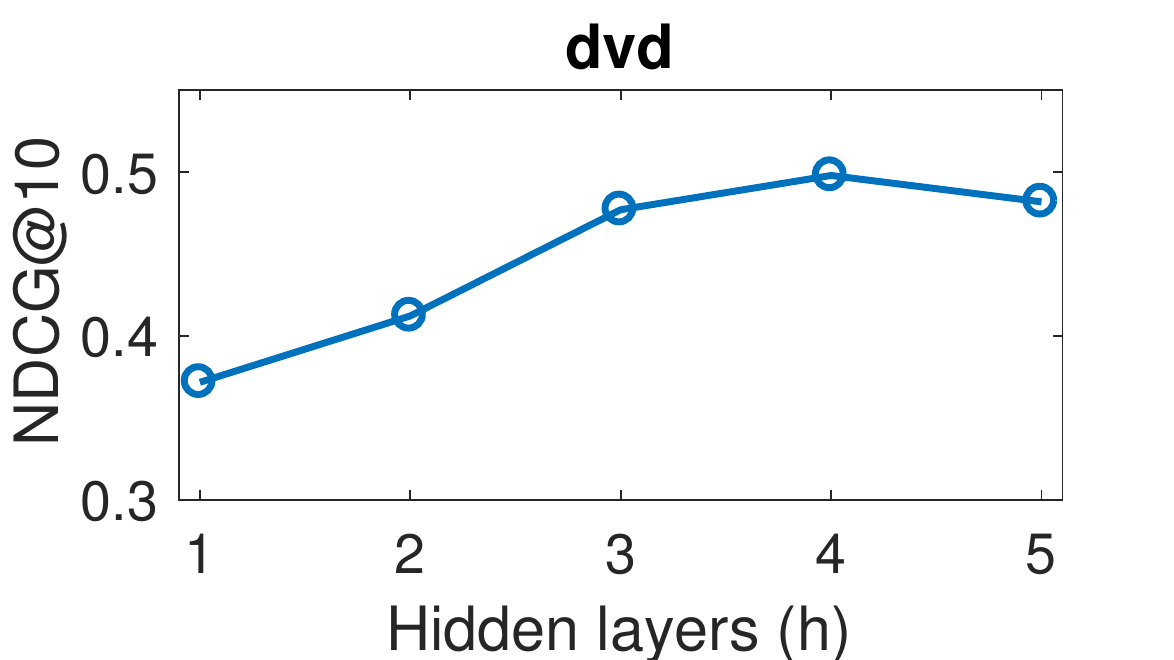}\\
 \includegraphics[width=\columnwidth]{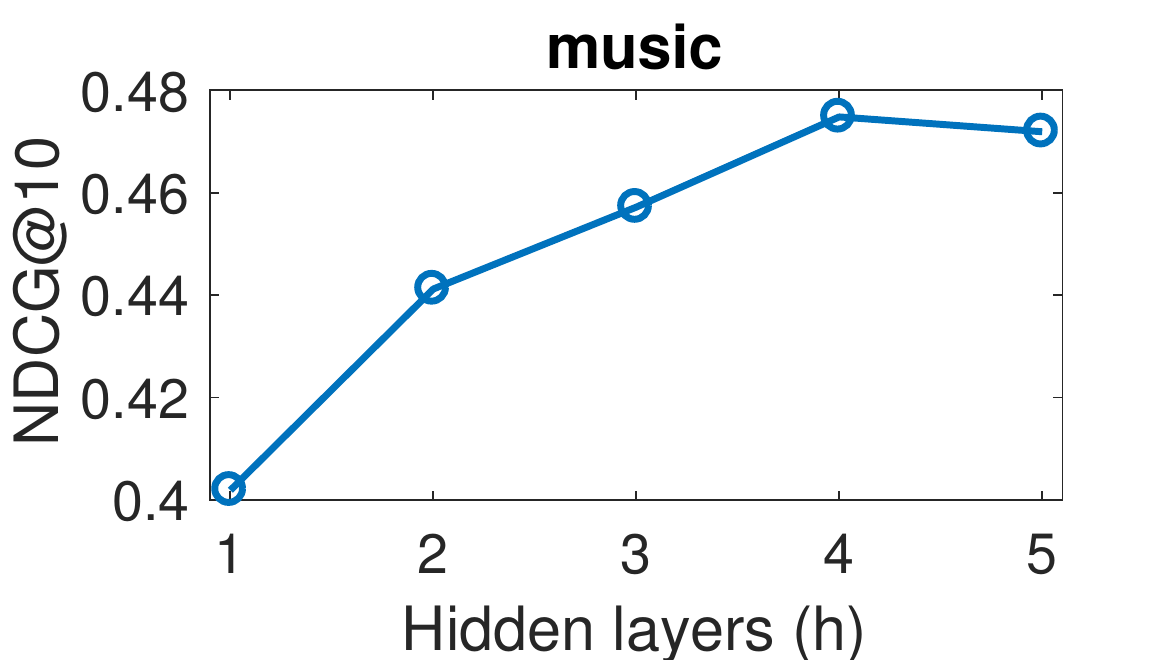} \hspace{0.1cm} &  \includegraphics[width=\columnwidth]{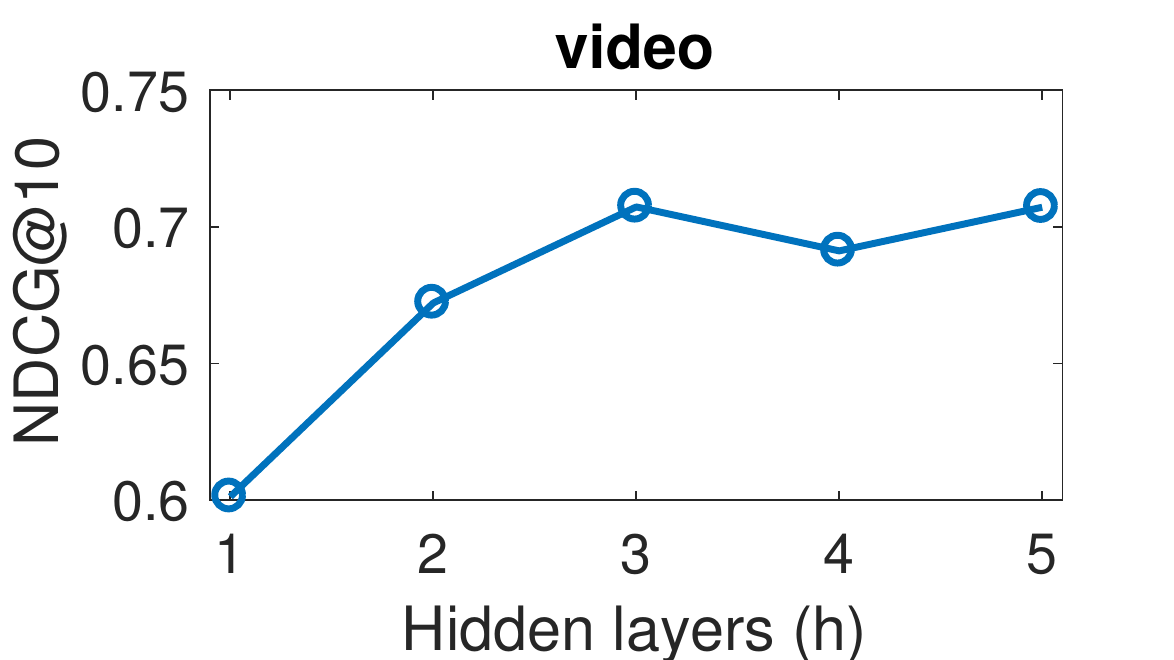} \\
\end{tabular}
\vspace{-0.1cm}\caption{Effect on NDCG when varying the number of shared hidden layers $h$ in our cross-domain neural architecture of Figure~\ref{fig:over}.} \label{fig:res2}
\end{figure*}

\subsection{Parameter Analysis} \label{sec:param}
\textbf{The influence of the shared hidden layers.} In the proposed ADC model to better capture the non-linearity of user preferences in multiple domains, we use $h$ shared hidden layers as illustrated in our cross-domain neural architecture of Figure~\ref{fig:over}, in order to train the model parameters based on the cross-domain loss function via our adaptive learning strategy. The number of $h$ hidden shared layers is an important parameter of our ADC model, demonstrating the importance of deep learning when generating cross-domain recommendations. In this set of experiments, we fix the number of negative samples $|{\mathcal{I}^-_u}^{(k)}|$=5 for each observed rating/sample of each domain, and the number of latent dimensions $d$=100. Notice that we keep the number of negative samples and latent dimensions fixed as we observed that after the fixed values ADC did not significantly increase the recommendation accuracy without paying off in terms of computational cost.  Figure~\ref{fig:res2} presents the effect on NDCG when varying the number of hidden layers from 1 to 5 by a step of 1. On inspection of the results in Figure~\ref{fig:res2} we set $h$=2 for the smallest domain ``toys'', $h$=4 for the largest domains ``dvd'' and ``music'', while for the remaining domains ``electronics'', ``kitchen'' and ``video'' that have similar complexities/scales we set $h$=3. Clearly, the number of hidden layers depends on the complexity/scale of the target domain, where more hidden layers are required for larger target domains. 

\begin{figure*}[t] 
\begin{tabular}{cc} 
\includegraphics[width=\columnwidth]{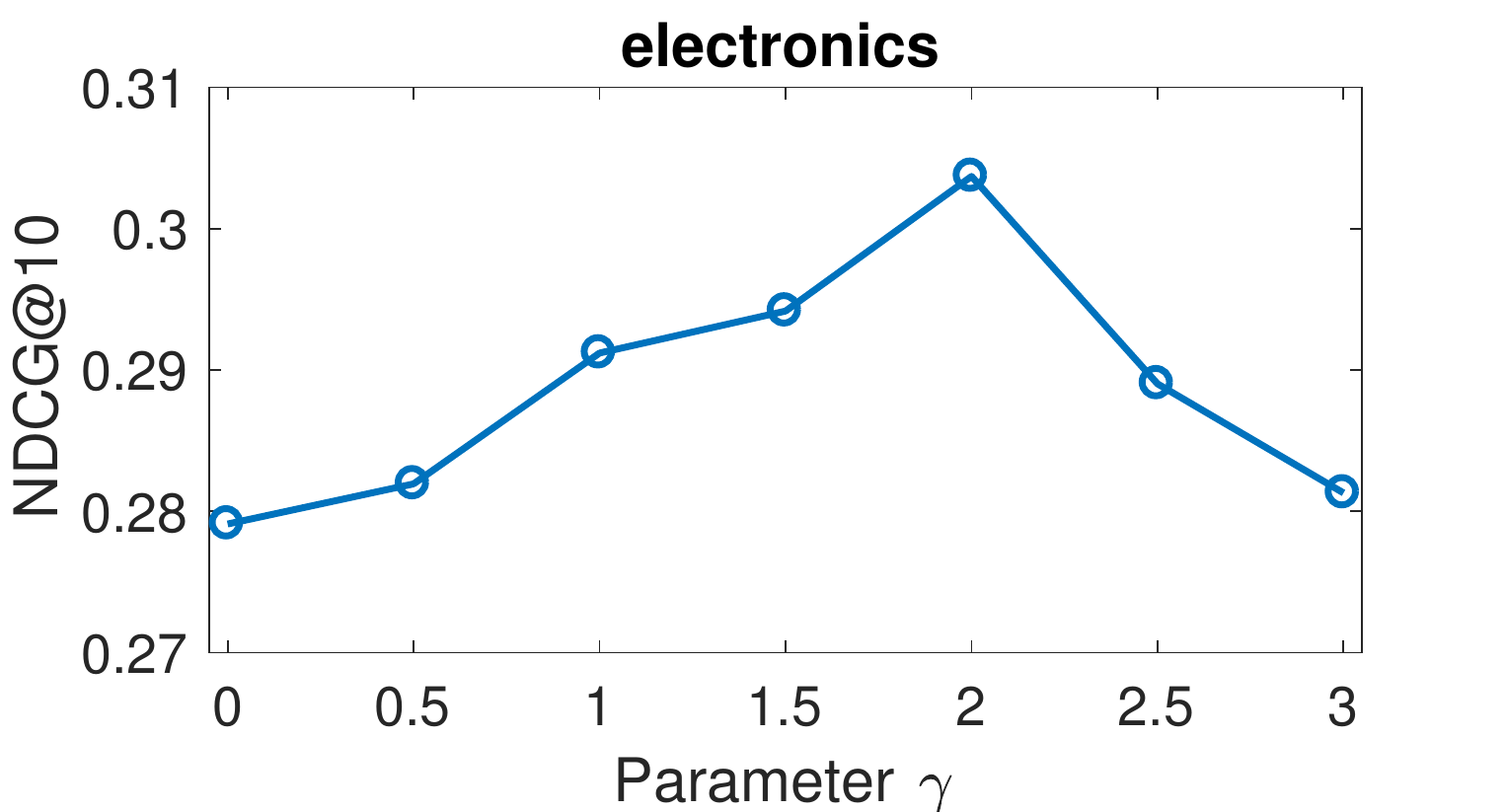} \hspace{0.1cm} & \includegraphics[width=\columnwidth]{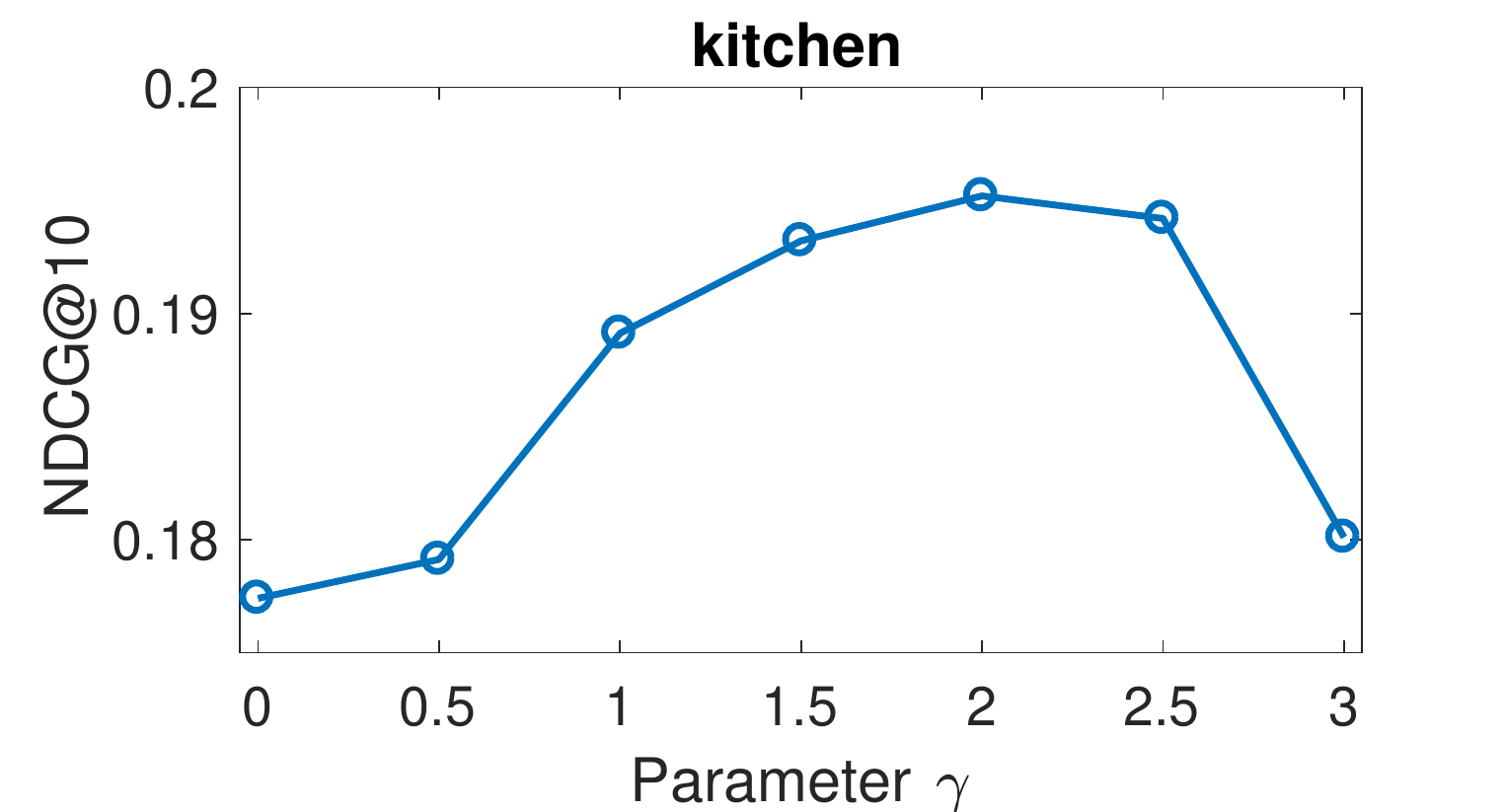} \\
\includegraphics[width=\columnwidth]{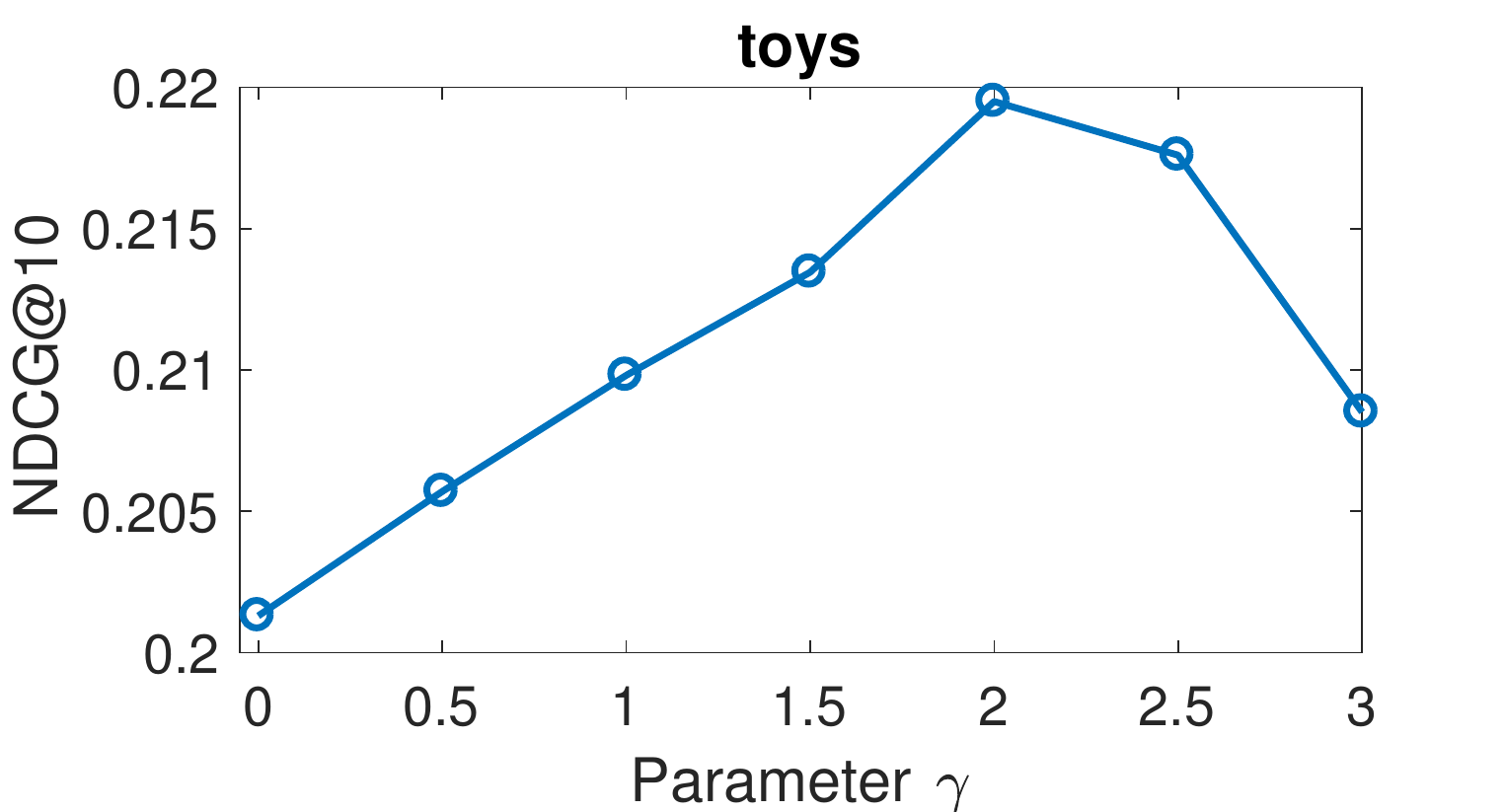}  \hspace{0.1cm} & \includegraphics[width=\columnwidth]{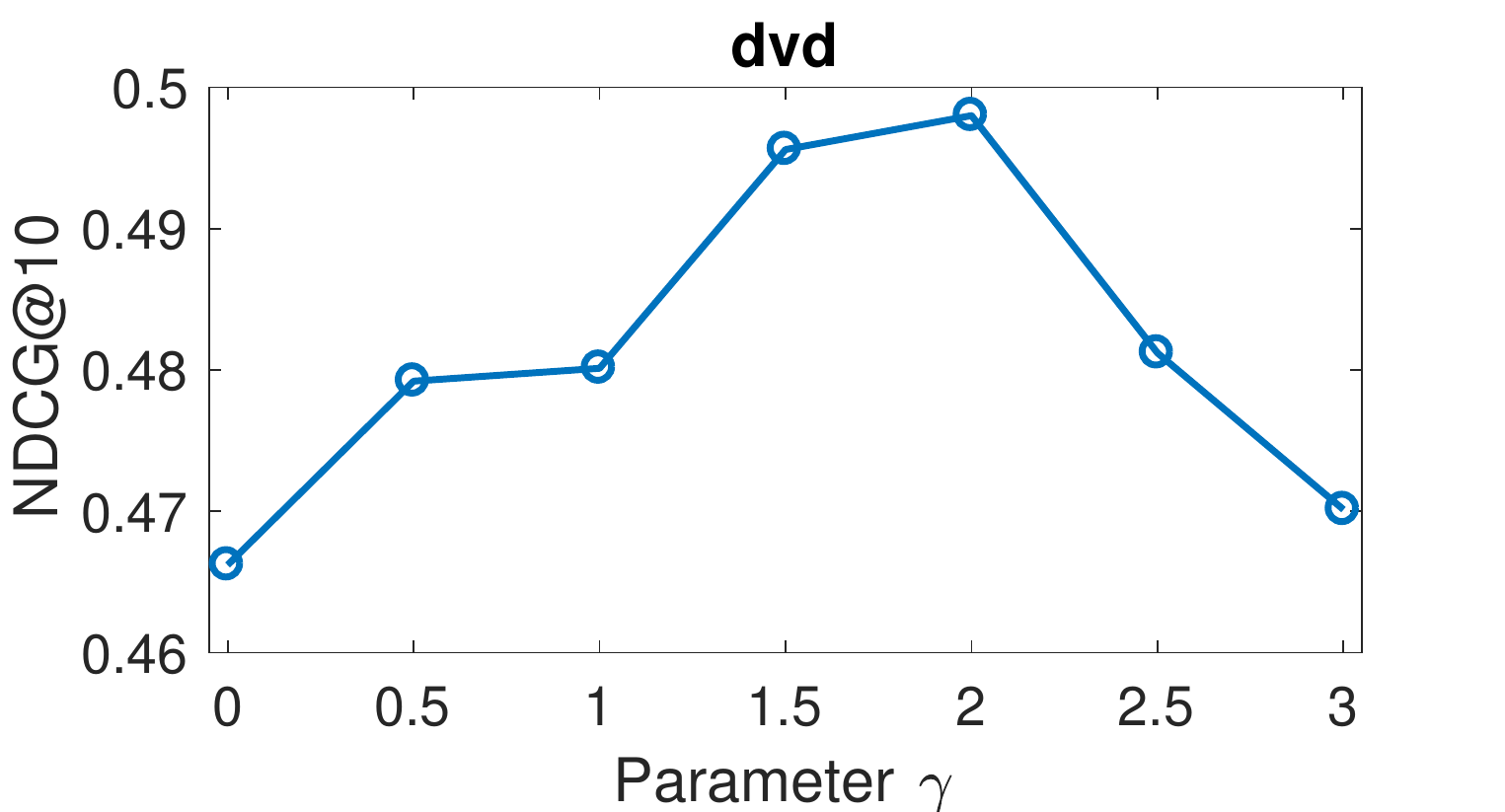}\\
 \includegraphics[width=\columnwidth]{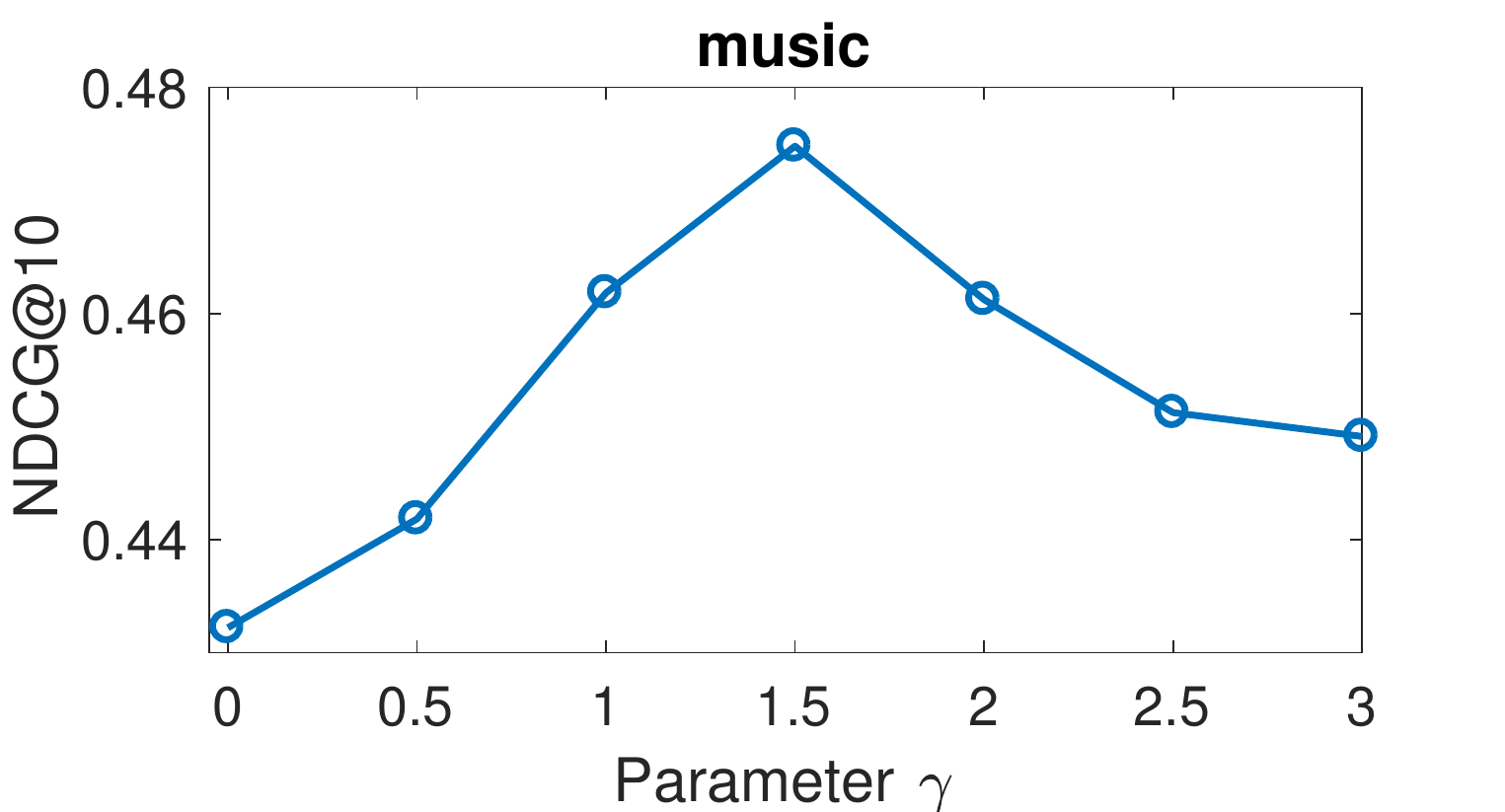} \hspace{0.1cm} &  \includegraphics[width=\columnwidth]{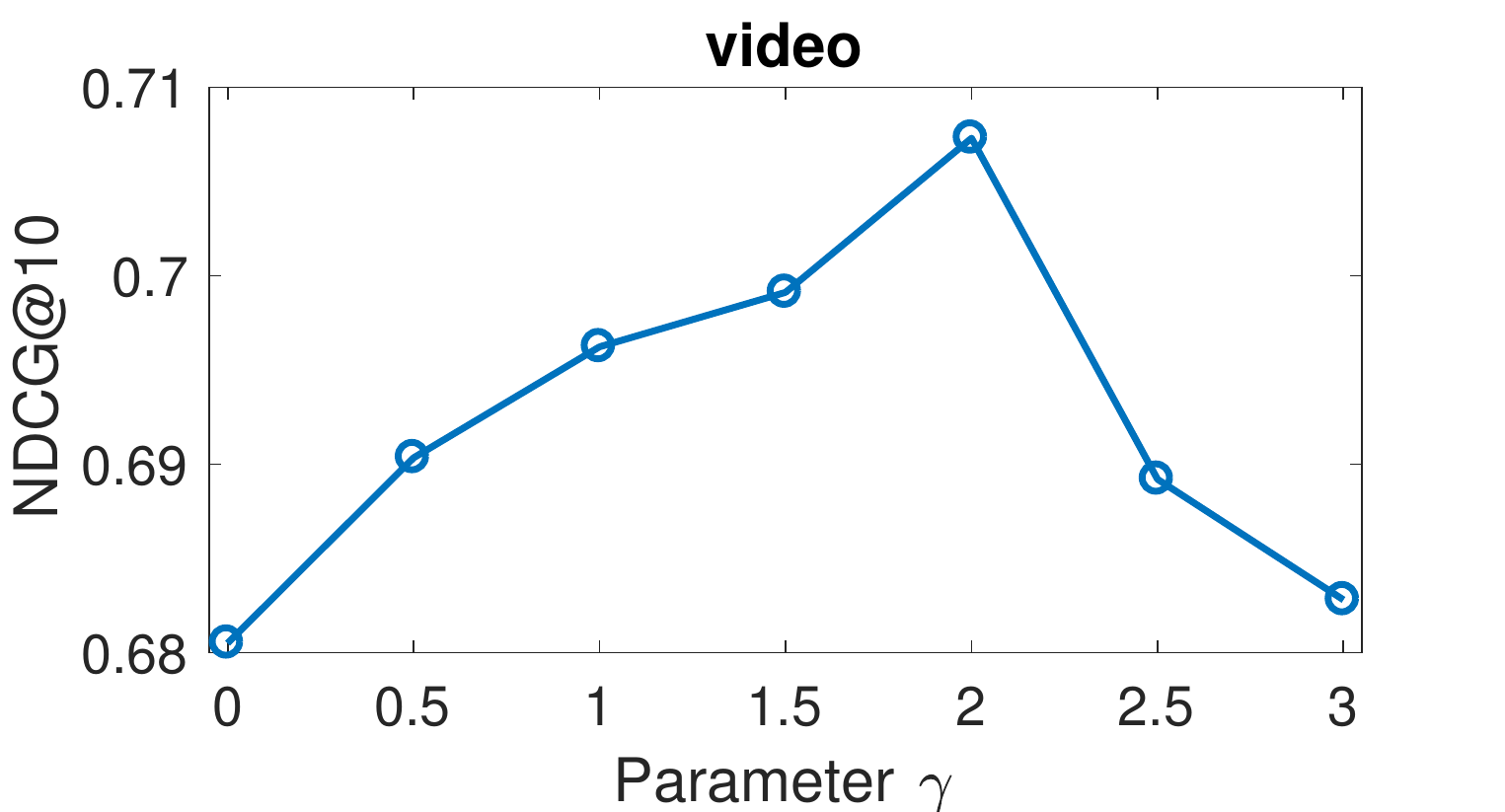} \\
\end{tabular}
\vspace{-0.1cm}\caption{Effect on NDCG when varying the \emph{asymmetry} parameter $\gamma$ in the loss function $L_{grad}$.} \label{fig:res3}
\end{figure*}

\textbf{The effect of the asymmetry parameter.} As stated in Section~\ref{sec:adap}, the hyperparameter $\gamma$ controls the adjustment of gradients magnitudes and learning rates for each domain in the loss function $L_{grad}$ of Equation~\ref{eq:tun} based on the domains' complexities/scales. In Figure~\ref{fig:res3} we examine the performance of the proposed ADC model when varying the hyperparameter $\gamma$ in [0, 0.5, 1, 1.5, 2, 2.5, 3]. Notice that lower $\gamma$ values are  required for more symmetric domains, while higher $\gamma$ values penalize the network when backpropagated gradients from any domain are too large or too small. The special case of $\gamma$= 0 corresponds to equal weights $w_k$=1 in the cross-domain loss function of Equation~\ref{eq:loss3}. Clearly, as we can observe from the results from Figure~\ref{fig:res3}, ADC under-performs when setting   $\gamma$ = 0. This occurs because of the asymmetry in the evaluation domains, as shown in Table~\ref{tab:data}, demonstrating that our adaptive learning strategy plays a key role in the performance of the proposed ADC model. Due to the asymmetric nature of the evaluation domains, we set $\gamma=2$ based on the experimental results of Figure~\ref{fig:res3} for all domains, as the model becomes unstable for different $\gamma$ values. In the exceptional case of the largest domain ``music'' we fix $\gamma=1.5$, as the ``music''  domain has the highest complexity, thus is less affected by the gradient and learning rate balancing of our adaptive learning strategy. 

To further evaluate the performance of our adaptive learning strategy we repeated our experiments by splitting the six domains of Table~\ref{tab:data} into the following two symmetric subsets of domains, that is, the two largest domains in Set 1=\{``dvd'', ``music''\} and the remaining domains in Set 2=\{``electronics'', ``kitchen'', ``toys'', ``video''\}. The best $\gamma$ values are 1 and 0.5 for Set 1 and Set 2, respectively. As expected, compared to the case of all six domains, less balancing (lower $\gamma$ value) in the loss function $L_{grad}$ is required, as both subsets Set 1 and Set 2 are more symmetric. Finally, to further verify the importance of our adaptive learning strategy, we repeated our experiments by generating an extremely asymmetric subset Set 3=\{``music'',  ``toys''\}, that is, the largest and smallest domains respectively. In this asymmetric case of Set 3 we fix $\gamma$=2.5, requiring more balancing (higher $\gamma$ value) compared to the case of all six domains. Our experimental results confirm that indeed an adaptive learning strategy is needed when generating cross-domain recommendation for domains with different complexities/scales.

\section{Conclusions}\label{sec:conc}
In this paper we presented ADC, an adaptive deep learning strategy for cross-domain recommendations. The two key factors of ADC are (i) to capture the non-linear associations of user preferences across domains by formulating a joint cross-domain loss function in our deep learning strategy and (ii) to adjust and weigh the influence of each domain when optimizing the model parameters based on the domains' complexities by applying an adaptive learning algorithm. Our experimental evaluation on six cross-domain recommendation tasks demonstrated the effectiveness of the proposed ADC model, evaluated against other state-of-the-art methods. Compared to the second best method, the proposed ADC model attains an average improvement of 6.14 and 8.08\%  in terms of recall and NDCG in all runs. In addition, we evaluated the performance of the proposed ADC model, when the adjustment of the weights in the cross-domain loss function is missing, demonstrating the importance of the proposed adaptive deep learning strategy. Furthermore, we studied the performance of ADC in the cases of both symmetric and asymmetric subsets of domains, and we observed that in both cases the proposed adaptive deep learning strategy has a stable performance, by adjusting the weights when optimizing the model parameters accordingly.

An interesting future direction is to exploit the proposed cross-domain balancing mechanism for social event detection~\cite{RafailidisSLSD13}, information diffusion~\cite{AntarisRN14,RafailidisN15}, exploiting social information~\cite{Rafailidis16a,sigir/RafailidisC16a} and capturing users' preference dynamics~\cite{RafailidisKM17,RafailidisN15a}. In addition, as future work, we plan to extend the proposed model for sequential recommendations in cross-domain tasks. Generating sequential recommendations is a challenging task, where the goal is to predict the next item that a user will select. Although there are many single-domain strategies for sequential recommendation, such as the studies reported in~\cite{Pas18,Quad17,Yuan19}, the case of cross-domain reflects better on the real-world scenario, where not only we have to capture users' sequential behaviors, but transfer this knowledge across different domains. 

\bibliographystyle{unsrt}
\bibliography{crossRec}

\begin{thebibliography}{10}

\bibitem{AliannejadiRC17}
Mohammad Aliannejadi, Dimitrios Rafailidis, and Fabio Crestani.
\newblock Personalized keyword boosting for venue suggestion based on multiple
  lbsns.
\newblock In {\em Advances in Information Retrieval - 39th European Conference
  on {IR} Research, {ECIR}, Aberdeen, UK, April 8-13}, pages 291--303, 2017.

\bibitem{RafailidisC16}
Dimitrios Rafailidis and Fabio Crestani.
\newblock Top-n recommendation via joint cross-domain user clustering and
  similarity learning.
\newblock In {\em Machine Learning and Knowledge Discovery in Databases -
  European Conference, {ECML} {PKDD}, Riva del Garda, Italy, September 19-23
  Proceedings, Part {II}}, pages 426--441, 2016.

\bibitem{RafailidisC17}
Dimitrios Rafailidis and Fabio Crestani.
\newblock A collaborative ranking model for cross-domain recommendations.
\newblock In {\em Proceedings of the {ACM} International Conference on
  Information and Knowledge Management, {CIKM}, Singapore, November 06 - 10},
  pages 2263--2266, 2017.

\bibitem{CREM10}
Paolo Cremonesi, Yehuda Koren, and Roberto Turrin.
\newblock Performance of recommender algorithms on top-n recommendation tasks.
\newblock In {\em Proceedings of the ACM Conference on Recommender Systems
  {RecSys}}, pages 39--46, 2010.

\bibitem{Her04}
Jonathan~L. Herlocker, Joseph~A. Konstan, Loren~G. Terveen, and John Riedl.
\newblock Evaluating collaborative filtering recommender systems.
\newblock {\em {ACM} Transaction on Information Systems}, 22(1):5--53, 2004.

\bibitem{Kor09}
Yehuda Koren, Robert~M. Bell, and Chris Volinsky.
\newblock Matrix factorization techniques for recommender systems.
\newblock {\em {IEEE} Computer}, 42(8):30--37, 2009.

\bibitem{RafailidisNM11}
Dimitrios Rafailidis, Alexandros Nanopoulos, and Yannis Manolopoulos.
\newblock Nonlinear dimensionality reduction for efficient and effective audio
  similarity searching.
\newblock {\em Multimedia Tools Appl.}, 51(3):881--895, 2011.

\bibitem{sigir/RafailidisC16}
Dimitrios Rafailidis and Fabio Crestani.
\newblock Cluster-based joint matrix factorization hashing for cross-modal
  retrieval.
\newblock In {\em Proceedings of the 39th International {ACM} {SIGIR}
  conference on Research and Development in Information Retrieval, {SIGIR},
  Pisa, Italy, July 17-21}, pages 781--784, 2016.

\bibitem{LI09}
Bin Li, Qiang Yang, and Xiangyang Xue.
\newblock Can movies and books collaborate? cross-domain collaborative
  filtering for sparsity reduction.
\newblock In {\em Proceedings of the International Joint Conference on
  Artificial Intelligence, {IJCAI}}, pages 2052--2057, 2009.

\bibitem{CREM11}
Paolo Cremonesi, Antonio Tripodi, and Roberto Turrin.
\newblock Cross-domain recommender systems.
\newblock In {\em Proceedings of the {IEEE} International Conferencecon Data
  Mining Workshops (ICDMW)}, pages 496--503, 2011.

\bibitem{TKDD15}
Nima Mirbakhsh and Charles~X. Ling.
\newblock Improving top-n recommendation for cold-start users via cross-domain
  information.
\newblock {\em {ACM} Transactions on Knowledge Discovery from Data},
  9(4):33:1--33:19, 2015.

\bibitem{Khan17}
Muhammad~Murad Khan, Roliana Ibrahim, and Imran Ghani.
\newblock Cross domain recommender systems: A systematic literature review.
\newblock {\em ACM Comput. Surv.}, 50(3):36:1--36:34, 2017.

\bibitem{BER07}
Shlomo Berkovsky, Tsvi Kuflik, and Francesco Ricci.
\newblock Distributed collaborative filtering with domain specialization.
\newblock In {\em Proceedings of the ACM Conference on Recommender Systems
  RecSys}, pages 33--40, 2007.

\bibitem{GAO13}
Sheng Gao, Hao Luo, Da~Chen, Shantao Li, Patrick Gallinari, and Jun Guo.
\newblock Cross-domain recommendation via cluster-level latent factor model.
\newblock In {\em Proceedings of the European Conference on Machine Learning
  and Knowledge Discovery in Databases, {ECML} {PKDD}, Part {II}}, pages
  161--176, 2013.

\bibitem{PAN10}
Weike Pan, Evan~Wei Xiang, Nathan~Nan Liu, and Qiang Yang.
\newblock Transfer learning in collaborative filtering for sparsity reduction.
\newblock In {\em Proceedings of the {AAAI} Conference on Artificial
  Intelligence, {AAAI}}, 2010.

\bibitem{HU13}
Liang Hu, Jian Cao, Guandong Xu, Longbing Cao, Zhiping Gu, and Can Zhu.
\newblock Personalized recommendation via cross-domain triadic factorization.
\newblock In {\em Proceedings of the {ACM} International Conference on World
  Wide Web, {WWW}}, pages 595--606, 2013.

\bibitem{Lon14}
Babak Loni, Yue Shi, Martha Larson, and Alan Hanjalic.
\newblock Cross-domain collaborative filtering with factorization machines.
\newblock In {\em Proceedings of the European Conference on Information
  Retrieval, {ECIR}}, pages 656--661, 2014.

\bibitem{EL15}
Ali~Mamdouh Elkahky, Yang Song, and Xiaodong He.
\newblock A multi-view deep learning approach for cross domain user modeling in
  recommendation systems.
\newblock In {\em Proceedings of the {ACM} International Conference on World
  Wide Web, {WWW}}, pages 278--288, 2015.

\bibitem{Hu18}
Guangneng Hu, Yu~Zhang, and Qiang Yang.
\newblock Conet: Collaborative cross networks for cross-domain recommendation.
\newblock In {\em Proceedings of the {ACM} International Conference on
  Information and Knowledge Management, {CIKM}}, pages 667--676, 2018.

\bibitem{Wang19}
Hongwei Wang, Fuzheng Zhang, Miao Zhao, Wenjie Li, Xing Xie, and Minyi Guo.
\newblock Multi-task feature learning for knowledge graph enhanced
  recommendation.
\newblock {\em CoRR}, abs/1901.08907, 2019.

\bibitem{Has17}
Kazuma Hashimoto, Caiming Xiong, Yoshimasa Tsuruoka, and Richard Socher.
\newblock A joint many-task model: Growing a neural network for multiple {NLP}
  tasks.
\newblock In {\em Proceedings of the 2017 Conference on Empirical Methods in
  Natural Language Processing, {EMNLP}}, pages 1923--1933, 2017.

\bibitem{Wu15}
Zhizheng Wu, Cassia Valentini{-}Botinhao, Oliver Watts, and Simon King.
\newblock Deep neural networks employing multi-task learning and stacked
  bottleneck features for speech synthesis.
\newblock In {\em Proceedings of the {IEEE} International Conference on
  Acoustics, Speech and Signal Processing, {ICASSP}}, pages 4460--4464, 2015.

\bibitem{Heg17}
Kaiming He, Georgia Gkioxari, Piotr Doll{\'{a}}r, and Ross~B. Girshick.
\newblock Mask {R-CNN}.
\newblock In {\em Proceedings of the {IEEE} International Conference on
  Computer Vision, {ICCV}}, pages 2980--2988, 2017.

\bibitem{Red17}
Joseph Redmon and Ali Farhadi.
\newblock {YOLO9000:} better, faster, stronger.
\newblock In {\em Proceedings of the {IEEE} Conference on Computer Vision and
  Pattern Recognition, {CVPR}}, pages 6517--6525, 2017.

\bibitem{Ken18}
Alex Kendall, Yarin Gal, and Roberto Cipolla.
\newblock Multi-task learning using uncertainty to weigh losses for scene
  geometry and semantics.
\newblock In {\em Proceeding ot the {IEEE} Conference on Computer Vision and
  Pattern Recognition, {CVPR}}, pages 7482--7491, 2018.

\bibitem{Mis16}
Ishan Misra, Abhinav Shrivastava, Abhinav Gupta, and Martial Hebert.
\newblock Cross-stitch networks for multi-task learning.
\newblock In {\em Proceedings of the {IEEE} Conference on Computer Vision and
  Pattern Recognition, {CVPR}}, pages 3994--4003, 2016.

\bibitem{Rendle12}
Steffen Rendle.
\newblock Factorization machines with libfm.
\newblock {\em {ACM} Transactions on Intelligent Systems and Technology},
  3(3):57:1--57:22, 2012.

\bibitem{BPR}
Steffen Rendle, Christoph Freudenthaler, Zeno Gantner, and Lars Schmidt-Thieme.
\newblock Bpr: Bayesian personalized ranking from implicit feedback.
\newblock In {\em Proceedings of the Conference on Uncertainty in Artificial
  Intelligence, {UAI}}, pages 452--461, 2009.

\bibitem{Les07}
Jure Leskovec, Lada~A. Adamic, and Bernardo~A. Huberman.
\newblock The dynamics of viral marketing.
\newblock {\em ACM Transactions on the Web}, 1(1):5, 2007.

\bibitem{Rec17}
Dimitrios Rafailidis and Fabio Crestani.
\newblock Learning to rank with trust and distrust in recommender systems.
\newblock In {\em Proceedings of the {ACM} Conference on Recommender Systems,
  RecSys}, pages 5--13, 2017.

\bibitem{RafailidisSLSD13}
Dimitrios Rafailidis, Theodoros Semertzidis, Michalis Lazaridis, Michael~G.
  Strintzis, and Petros Daras.
\newblock A data-driven approach for social event detection.
\newblock In {\em Proceedings of the MediaEval 2013 Multimedia Benchmark
  Workshop, Barcelona, Spain, October 18-19}, 2013.

\bibitem{AntarisRN14}
Stefanos Antaris, Dimitrios Rafailidis, and Alexandros Nanopoulos.
\newblock Link injection for boosting information spread in social networks.
\newblock {\em Social Netw. Analys. Mining}, 4(1):236, 2014.

\bibitem{RafailidisN15}
Dimitrios Rafailidis and Alexandros Nanopoulos.
\newblock Crossing the boundaries of communities via limited link injection for
  information diffusion in social networks.
\newblock In {\em Proceedings of the 24th International Conference on World
  Wide Web Companion, {WWW} 2015, Florence, Italy, May 18-22, - Companion
  Volume}, pages 97--98, 2015.

\bibitem{Rafailidis16a}
Dimitrios Rafailidis.
\newblock Modeling trust and distrust information in recommender systems via
  joint matrix factorization with signed graphs.
\newblock In {\em Proceedings of the 31st Annual {ACM} Symposium on Applied
  Computing, Pisa, Italy, April 4-8}, pages 1060--1065, 2016.

\bibitem{sigir/RafailidisC16a}
Dimitrios Rafailidis and Fabio Crestani.
\newblock Collaborative ranking with social relationships for top-n
  recommendations.
\newblock In {\em Proceedings of the 39th International {ACM} {SIGIR}
  conference on Research and Development in Information Retrieval, {SIGIR},
  Pisa, Italy, July 17-21}, pages 785--788, 2016.

\bibitem{RafailidisKM17}
Dimitrios Rafailidis, Pavlos Kefalas, and Yannis Manolopoulos.
\newblock Preference dynamics with multimodal user-item interactions in social
  media recommendation.
\newblock {\em Expert Syst. Appl.}, 74:11--18, 2017.

\bibitem{RafailidisN15a}
Dimitrios Rafailidis and Alexandros Nanopoulos.
\newblock Repeat consumption recommendation based on users preference dynamics
  and side information.
\newblock In {\em Proceedings of the 24th International Conference on World
  Wide Web Companion, {WWW} 2015, Florence, Italy, May 18-22 - Companion
  Volume}, pages 99--100, 2015.

\bibitem{Pas18}
Rajiv Pasricha and Julian McAuley.
\newblock Translation-based factorization machines for sequential
  recommendation.
\newblock In {\em Proceedings of the {ACM} Conference on Recommender Systems,
  {RecSys}}, pages 63--71, 2018.

\bibitem{Quad17}
Massimo Quadrana, Alexandros Karatzoglou, Bal{\'{a}}zs Hidasi, and Paolo
  Cremonesi.
\newblock Personalizing session-based recommendations with hierarchical
  recurrent neural networks.
\newblock In {\em Proceedings of the {ACM} Conference on Recommender Systems,
  {RecSys}}, pages 130--137, 2017.

\bibitem{Yuan19}
Fajie Yuan, Alexandros Karatzoglou, Ioannis Arapakis, Joemon~M. Jose, and
  Xiangnan He.
\newblock A simple convolutional generative network for next item
  recommendation.
\newblock In {\em Proceedings of the {ACM} International Conference on Web
  Search and Data Mining, {WSDM}}, pages 582--590, 2019.

\end{thebibliography}

\end{document}